\documentclass[aps,onecolumn,preprint,prd,superscriptaddress,nofootinbib]{revtex4}%
\pdfoutput=1
\usepackage{color,graphicx,epsfig}
\usepackage{ifpdf}
\usepackage{amsmath}
\usepackage{bm}
\usepackage{color}
\usepackage[english]{babel}
\usepackage{graphicx}%
\usepackage{amsfonts}%
\usepackage{amssymb}
\usepackage{braket}
\usepackage{hyperref}

\definecolor{nicered}{rgb}{0.7,0.1,0.1}
\definecolor{nicegreen}{rgb}{0.1,0.5,0.1}
\definecolor{niceblue}{rgb}{0.1,0.1,0.7}
\hypersetup{colorlinks,citecolor= nicegreen,linkcolor= niceblue}

\newcommand{\slashed}{\slash \hspace{-0.25cm}}

\newcommand{\cO}[0]{{\cal O}}
\newcommand{\cV}[0]{{\cal V}}
\newcommand{\beq}{\begin{equation}}
\newcommand{\eeq}{\end{equation}}
\newcommand{\bea}{\begin{eqnarray}}
\newcommand{\eea}{\end{eqnarray}}

\definecolor{Red}{rgb}{1.,0.,0.}

\preprint{CERN-TH-2018-180}

\begin{document}

\def\LjubljanaFMF{Faculty of Mathematics and Physics,\\ University of Ljubljana,
 Jadranska 19, 1000 Ljubljana, Slovenia }
\def\LjubljanaIJS{Jo\v zef Stefan Institute, Jamova 39, 1000 Ljubljana, Slovenia}
\def\CERN{Theoretical Physics Department, CERN, Geneva, Switzerland}
\def\UniGeneve{ D\'epartement de Physique Th\'eorique  
	and Center for Astroparticle Physics (CAP), Universit\'e de 
	Gen\`eve, 24 quai E. Ansermet, CH-1211, Geneva 4, Switzerland}
\def\Ottawa{Ottawa-Carleton Institute for Physics, Carleton University 1125 
	Colonel By Drive, Ottawa, Ontario K1S 5B6, Canada}

\title{On Lepton Flavor Universality in Top Quark Decays}

\author{Jernej F.\ Kamenik} 
\email[Electronic address:]{jernej.kamenik@ijs.si} 
\affiliation{\LjubljanaIJS}
\affiliation{\LjubljanaFMF}

\author{Andrey Katz} 
\email[Electronic address:]{andrey.katz@cern.ch} 
\affiliation{\CERN}
\affiliation{\UniGeneve}

\author{Daniel Stolarski} 
\email[Electronic address:]{stolar@physics.carleton.ca} 
\affiliation{\Ottawa\vspace{1cm}}


\begin{abstract}
\vspace{.3cm}
\noindent
We propose a novel strategy to test lepton flavor universality (LFU) in top decays, applicable to top pair production at 
colliders. Our proposal exploits information in kinematic distributions and mostly hinges on data-driven techniques, thus having very little dependence on our theoretical understanding of top pair production.
Based on simplified models accommodating recent hints of LFU violation in charged current B meson decays, we show that existing 
LHC measurements already provide non-trivial information on the flavor structure and the mass scale of such new
 physics (NP). We also project that the measurements of LFU in top decays at the {high-luminosity LHC} could reach a precision at the percent level or below, improving the sensitivity to LFU violating NP in the top sector by more than an order of magnitude compared to existing approaches. 


\end{abstract}

\maketitle
\tableofcontents
\section{Introduction}
\label{sec:intro}

Lepton flavor universality (LFU) of weak interactions is one of the key predictions of the Standard Model (SM). It has been well tested directly in on-shell $W$ and $Z$ decays at LEP~\cite{Patrignani:2016xqp,ALEPH:2005ab}.
Pion, kaon, charm and tau decays have also been studied extensively in the past, confirming LFU to a precision ranging between a percent and a per-mille level~\cite{Patrignani:2016xqp}.  Recently, however, 
hints of violation of LFU at the level of 20\% have appeared in measurements of charged current mediated semi-tauonic B meson decays~\cite{Amhis:2016xyh} as well as in rare flavor changing neutral current mediated semi-muonic and semi-electronic B meson decays~\cite{Aaij:2014ora,Aaij:2015oid,Aaij:2017vbb}. These hints might indicate beyond SM contributions to weak interactions of third and second generation SM fermions (see e.g. Refs.~\cite{Fajfer:2012jt,Freytsis:2015qca, Bardhan:2016uhr, Celis:2016azn, Bernlochner:2017jka} and~\cite{Hiller:2014yaa, Altmannshofer:2014rta, Altmannshofer:2015sma,Descotes-Genon:2015uva, Capdevila:2017bsm} for  some general 
analyses). On the one hand it is imperative to verify the intriguing results in B decays with additional observables both involving $b$-hadrons, but as well in other flavor sectors of the theory. More generally LFU observables can be predicted with high accuracy within the SM and are typically also very clean experimentally. Thus they form key tests of the SM flavor sector and also important vectors in the search for hints of possible new physics (NP) indirectly. 

LFU in top decays is currently much less established experimentally. The precision of existing LHC measurements of the top decay branching fractions to final states involving a $\tau$ lepton is at the $20\%$ level and already limited by systematic uncertainties~\cite{Aad:2015dya,CMS-PAS-HIG-12-052}. Despite this relatively poor precision, we find that these searches have comparable sensitivity to new physics for mediators lighter than the top mass. If, on the other hand, the new mediators are heavier than the top mass, the effects in top physics become significantly smaller.

In the present work we propose a novel experimental strategy that can improve the sensitivity of top decay measurements at the LHC to the presence of possible LFU violating NP by more than an order of magnitude. The key insight is that heavy (off-shell) NP contributions to $t \to b \tau \nu$
 decays  will result in final state distributions distinctly different from the SM two-body $t \to W b$ kinematics. This can be used to probe tiny LFU violating effects in localized regions of phase-space which might be averaged out in the integrated total rate measurements. In addition, we propose several data-driven techniques in order to leverage sufficient control over possible systematics effects. 
As we will show, this opens up the possibility to probe sub-percent level LFU violating effects at the (HL) LHC.

The remainder of the paper is structured as follows: in Secs.~\ref{sec:eft} and~\ref{sec:simplified} we introduce examples of possible LFU violating NP affecting charged current top decays both in  the language of effective field theory, and in terms of simplified models, respectively. In Sec.~\ref{sec:bounds} we review the existing measurements of LFU in top decays and derive the corresponding constraints on our NP examples. Sec.~\ref{sec:idea} contains a basic introduction to our search strategy and a demonstration of its discriminating power against possible LFU violating NP at parton level. This is followed in Sec.~\ref{sec:details} by the description of the explicit implementation of our method including a detailed discussion of the related systematic uncertainties and control thereof. Finally, a recap of our main results and our conclusions are presented in Sec.~\ref{sec:conclusions}. Some analytic results regarding the NP effects 
on the b-quarks energy peak are relegated to the appendix.

\section{Effective field theory of LFU violation in top and B physics}
\label{sec:eft}



We start our discussion of possible LFU effects in top physics in an effective field theory (EFT) language, suitable for phenomenological studies in presence of heavy NP. In particular, provided new degrees of freedom are much heavier than the energy scales relevant to top decays, one can describe the most general departures from the SM predictions in charged current (semi) tauonic top quark transitions in terms of only a few effective operator structures\footnote{In the following we do 
	not consider the possibility that the missing energy signature of SM neutrinos in weak decays is mimicked by the presence of new light neutral particles. See however the related discussion in Refs.~\cite{Fajfer:2012jt,Becirevic:2016yqi}.} appearing at the lowest operator dimension (six)~\cite{Freytsis:2015qca}. 
Below the weak scale one can thus describe the relevant EFT including the leading NP LFU violating contributions as
\beq
\mathcal L_{\rm EFT} = \mathcal L_{\rm SM} + \frac{1}{\Lambda^2}\sum_{i,q} C^q_i \mathcal O^q_i + \rm h.c.\,,
\eeq
where $\Lambda$ is the EFT cut-off (or matching) scale, $C^q_i$ are the relevant Wilson coefficients and $\mathcal O^q_i$ the corresponding EFT operators involving a quark of flavor $q$. For simplicity we assume $C^q_i$ to be real. In the following we restrict our discussion to operators, which (1) can be related (either via the SM $SU(2)_L$ gauge invariance or through rotations in quark flavor space) to operators mediating semitauonic B-meson weak decays, and (2) can be most easily matched in the UV to well defined simplified models -- SM extensions with a single new field in some SM gauge representation.  Only few particular combinations of $\mathcal O^q_i$ satisfy both criteria~\cite{Freytsis:2015qca}. 
Defining the set of operators
\begin{align}
\mathcal O^q_{VL} & = (\bar q \gamma_\mu P_L b ) (\bar \tau \gamma^\mu P_L \nu_\tau)\,, & \hskip-0.1cm \mathcal O^q_{SL} & = (\bar q  P_L b ) (\bar \tau  P_L \nu_\tau), \nonumber\\
\mathcal O^q_{TL} & = (\bar q\sigma_{\mu\nu} P_L b ) (\bar \tau \sigma^{\mu\nu} P_L \nu_\tau)\,, &\hskip-0.1cm \mathcal O^q_{SR} & = (\bar q  P_R b ) (\bar \tau  P_L \nu_\tau) \,,
\end{align}
where $P_{R,L} \equiv (1\pm \gamma_5)/2$, the following parameter benchmark points have been found to reproduce the current experimental results and can be matched to well defined simplified models~\cite{Freytsis:2015qca}: (a) $\bar C^c_{VL} = 0.18 (4)$, 
(b) $\bar C^c_{SL} \simeq -1.02  $ and $\bar C^c_{SR} \simeq 1.25 $, and finally (c) $-2 \bar C^c_{SL} = 8 \bar C_{TL} = -0.46(9)$, where we have used a short-hand notation $\bar C^q_{i} \equiv  C^q_{i} ({1\rm TeV}/{\Lambda})^2$. While these values are chosen to reproduce recent observational hints for LFU violation in charged current B decays, they can be also reinterpreted as representative of the size of LFU violating NP within reach of current precision B decay measurements. 


In order to relate departures from LFU of weak charged current interactions in the bottom and top quark sectors one also needs to specify the quark flavor structure of NP.  In light of severe constraints on new sources of quark and lepton flavor violation coming from FCNC observables and CKM unitarity tests (see e.g. Ref.~\cite{Alpigiani:2017lpj}), it is prudent to assume CKM-like hierarchies between the strengths of the various $b\leftrightarrow q$ flavor conversions, where $q=u,c,t$\,. In particular we employ $C_i^c / C_i^t  =   V_{cb} / V_{tb}$, where $V_{qb}$ are the relevant CKM elements. Relaxing this assumption leads to a straightforward rescaling of our results relating top and B physics observables which we briefly discuss in the final section. Translating the B physics benchmarks to top decays we obtain the expected deviations in the $t\to b\tau\nu$ decay branching fraction ($\delta \mathcal B_\tau \equiv \mathcal B_\tau / \mathcal B_\tau^{\rm SM}-1$) as
\begin{subequations}
\begin{align}
\label{eq:dB1}
&& {\rm (a)} ~~ \delta \mathcal B_\tau & = 1.8 \times 10^{-5} \bar C^t_{VL}  + 2.0 \times 10^{-5} (\bar C^t_{VL})^2   \,, \\
&& {\rm (b)} ~~  \delta \mathcal B_\tau & = 5.1 \times 10^{-6} \left[   (\bar C^t_{SL})^2 + (\bar C^t_{SR})^2\right] \,, \\
&& {\rm (c)} ~~  \delta \mathcal B_\tau & = 5.1 \times 10^{-6}    (\bar C^t_{SL})^2 + 2.4 \times 10^{-4} (\bar C^t_{TL})^2 \,.
\label{eq:dB3}
\end{align}
\end{subequations}
We first note that, while a strict EFT power counting would require to truncate the expansion of the above expressions at leading order in  $\bar C^t_{i}$, keeping also $(\bar C^t_{i})^2$ terms simplifies matching to dynamical NP models defined below. 
Inserting the values of the Wilson coefficients preferred by B decay data and assuming CKM-like flavor structure of NP, we observe that the expected effects are tiny and will be extremely challenging to probe. Here we also emphasize that although the deviations, motivated by the B physics hints, imply 
$\cO (10^{-5})$ deviations from the SM predicted values, the current bounds are four orders of magnitude larger. Irrespective of their connections to $B$ physics, any significant improvement from the current $\cO (20\%)$ sensitivity detailed 
in Sec~\ref{sec:bounds} is clearly worth pursuing.

Furthermore, as we will see in Sec~\ref{sec:idea} the interference effects of the NP with the SM might play an important 
role in the techniques that we propose. 
However, in the cases (b) and (c), the linear (interference) terms are suppressed by the $\tau$ or $b$-quark masses 
and thus completely negligible. Even in case (a)  terms quadratic in $\bar C^t_{i}$ still dominate over interference 
effects for the currently preferred parameter values. The smallness of the linear terms in this case can be simply 
understood by considering the partially integrated decay width as a function of the leptonic invariant mass squared 
$d\Gamma / d m^2_{\tau\nu} $, where $m^2_{\tau\nu} = (p_\tau+p_\nu)^2$. In the SM the overwhelming contribution to 
the width comes from the $W$ pole near $m^2_{\tau\nu}  = m_W^2$. The NP EFT contributions on the other hand are 
analytic in $m^2_{\tau\nu} $. The interference terms then pick up a phase rotation of $\pi$ when integrating close the $W$ 
pole. Since numerically the W width is much smaller than its mass which is furthermore roughly half the top mass, 
the interference contributions to $d\Gamma / d m^2_{\tau\nu} $ of opposite signs when integrated above and below the $W$
mass squared are comparable in size and cancel to a large extent. 
 
\section{Simplified models of LFU violation in top decays}
\label{sec:simplified}

The  EFT description discussed above fails at the mass scale of NP ($\Lambda$) where it should be matched onto a dynamical model involving new degrees of freedom. If the higher dimension operators are generated at tree level, the matching implies the presence of new EM charged particles.
 Existing LEP bounds~\cite{Heister:2002ev,Achard:2003gt,Abbiendi:2008aa} then require
  $\Lambda \gtrsim 100$~GeV.  While this confirms the EFT treatment of the B decays as adequate, the same is not necessarily 
  true for the top decays. We thus introduce three simplified models 
  (containing few fields beyond the SM, not necessarily renormalizable) which can be matched onto the EFT benchmarks relevant for B physics. In particular Model (a) consists of a massive charged spin-1 field ($\cV^{-}$) with the relevant Lagrangian given by
\begin{align}
\mathcal L^{(a)} &= \mathcal L_{\rm SM} + \frac{1}{4} \cV^+_{\mu\nu}\cV^{-\mu\nu} -
 m_\cV^2 \cV^+_\mu \cV^{-\mu} \nonumber\\
& +[ g_b  \sum_q V_{qb} \bar q \slashed \cV^+ P_L b + g_\tau  \bar \tau \slashed \cV^- P_L \nu_\tau + {\rm h.c.} ]\,, 
\end{align} 
where $\cV^+ \equiv (\cV^-)^\dagger$ and $\cV^\pm_{\mu\nu} \equiv \partial_\mu \cV^\pm_\nu - \partial_\nu \cV^\pm_\mu$\,. The EFT tree level matching conditions are then simply $C^q_{VL}/\Lambda^2 = g_\tau g_b V_{qb} / m_\rho^2$ with all other $C^q_i=0$\,. Models of this type have been considered in Refs.~\cite{Greljo:2015mma, Boucenna:2016wpr, Boucenna:2016qad}. Model (b) instead consists of a charged scalar ($\phi^-$)
\begin{align}
\mathcal L^{(b)} &= \mathcal L_{\rm SM} + \partial_\mu \phi^+ \partial^\mu \phi^-  - m_\phi^2 \phi^+ \phi^- \nonumber\\
& \hskip-0.5cm + [\sum_q V_{qb} \phi^+ (y^L_\phi \bar q P_L b + y^R_\phi \bar q P_R b)  + y^\tau_\phi \phi^- \bar \tau P_L \nu_\tau + {\rm h.c.} ]\,, 
\end{align} 
where now $\phi^+ \equiv (\phi^-)^\dagger$ and the tree-level matching conditions read $C^q_{SL}/\Lambda^2 = y_\phi^L y_\phi^\tau V_{qb} / m_\phi^2$, $C^q_{RL}/\Lambda^2 = y_\phi^R y_\phi^\tau V_{qb} / m_\phi^2$ with all other $C^q_i=0$. 
Such dynamics typically appears in two Higgs doublet models and has been studied extensively (see e.g. Refs.~\cite{Celis:2012dk, Crivellin:2013wna}). Finally, benchmark point (c) can be matched onto models of leptoquarks~\cite{Dorsner:2016wpm}, as considered for example in Ref.~\cite{Bauer:2015knc}. These being colored particles they can be efficiently pair produced at hadron colliders if within kinematical reach leading in turn to existing bounds on their masses much above the top quark 
mass~\cite{Aaboud:2016qeg,CMS-PAS-B2G-16-028}.  Consequently we do not consider a 
dynamical model for~(c) but work within the EFT as defined in the previous section even when discussing top decays.

\section{Bounds on top LFU violation from current measurements}
\label{sec:bounds}

While no dedicated experimental tests of LFU have yet been performed using the Tevatron or especially the large existing LHC top quark datasets, the branching fractions of top decays to final states involving different lepton flavors have already been measured individually. The currently most precise determination yields~\cite{Aad:2015dya}
 \begin{align}
 \mathcal B_e &= 13.3 (4)(4)\%\,, &  \hskip-0.25cm  \mathcal B_\mu &= 13.4 (3)(5)\%\,, &   \hskip-0.25cm \mathcal B_{\tau_h}&= 7.0(3)(5) \%\,,
\label{eq:Bl}
 \end{align}
 where $\mathcal B_\ell \equiv \mathcal B (t\to b \ell E_{\rm miss}) $ and $E_{\rm miss}$ denotes missing energy carried away by neutrinos. 
 The values in the first (second) parentheses refer to statistical (systematic) 
 uncertainties. 
 The modes with light leptons include contributions also from intermediate leptonic $\tau$ decays, 
 while the $\tau_h$ mode only accounts for $\tau$'s identified from their hadronic decays. 
 All three modes are in  agreement with SM LFU expectations at the one sigma level. 
 Solving the coupled system we can conclude that currently LFU in top decays is 
 tested at the $5-10\%$ uncertainty level between the $e$ and $\mu$ flavors, and 
 $15-25\%$ between the $\tau$ and the light lepton flavors, depending on the correlations 
 of systematic uncertainties between the three 
 modes.\footnote{The upper/lower limits of the ranges are obtained by including the 
 	systematic uncertainties in the measurements as uncorrelated or completely 
 	correlated, respectively.} 
 
\begin{figure}[!t]
\centering
\includegraphics[angle=0,width=10cm]{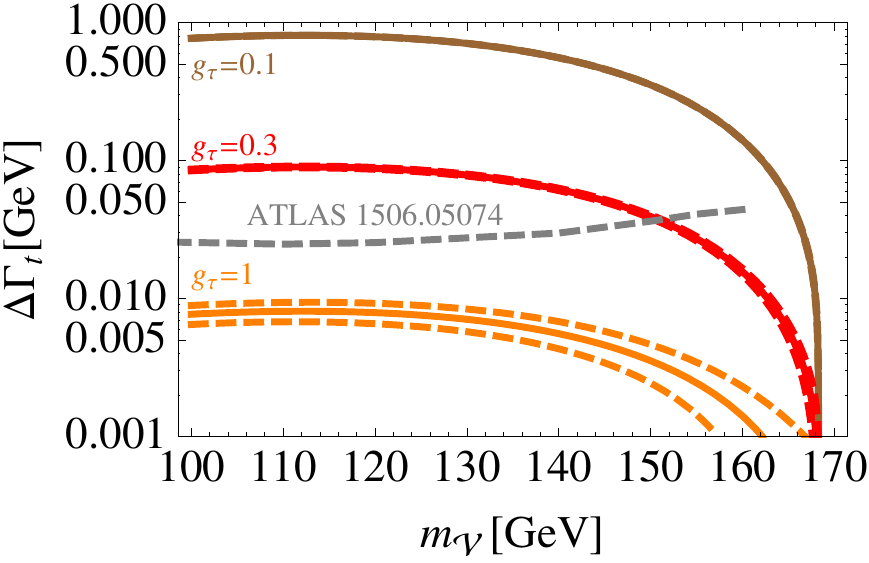}
\caption{\label{fig:1} The NP contribution to the total top width  ($\Delta \Gamma_t$)in the simplified Model (a). For each pair of values of 
$g_\tau$ and $m_{\cV}$ we choose the corresponding value of $g_b$ to accommodate the current B physics results, with the dashed lines indicating the $1\sigma$ band 
around the central values (solid lines). The grey dashed line indicates the 
ATLAS constraint from the top quark branching ratios measurements~\cite{Aad:2015dya}. See text for details. }
\end{figure}

 Unfortunately, since these measurements assume SM kinematics in top decays, 
 in particular the chain $t \to b W, W \to \ell\nu$, their results cannot be directly 
 applied to NP models. To estimate the sensitivity of such measurements to NP 
 contributions we recast the measurement including contributions of simplified Model~(a). 
 The chiral structure of interactions in this model 
 is identical to SM and the experimental signatures coincide exactly in 
 the limit $m_\cV = m_W$. We simulate the 
 NP signal and the SM events using \verb!MadGraph5_aMC@NLO!~\cite{Alwall:2014hca} and 
 {\tt Feynrules 2}~\cite{Alloul:2013bka} implementation of the model. 
 After {\tt Pythia~8}~\cite{Sjostrand:2014zea} 
 showering and hadronization we employ {\tt Delphes 3}~\cite{deFavereau:2013fsa} for fast detector simulation 
 and impose selection and isolation cuts matching those of Ref.~\cite{Aad:2015dya} for the various signal categories. 
 In the SM case we obtain reasonable agreement with the reported acceptance times efficiency ($(\epsilon \mathcal A)_{\rm SM}$) values of Ref.~\cite{Aad:2015dya}. We then use the ratio$(\epsilon \mathcal A)_{\rm NP}/ (\epsilon \mathcal A)_{\rm SM}$  to estimate the relative efficiency and acceptance corrections due to the different NP kinematics. 
 We find that these corrections range between $10\%$ at $m_\cV=100$~GeV, to 
 $50\%$ at $m_{\cV}=160$~GeV reducing the sensitivity to larger $\cV$ masses. 
 Since in this model $\mathcal B(\cV \to \tau\nu)\simeq 1$, 
 mostly $\mathcal B_{\tau_h}$ in Eq.~\eqref{eq:Bl} 
 is affected and we use this measurement to constrain the relevant parameter space. 
 After fixing the effective $\bar C^b_{VL}$ to the value allowed/preferred by B physics and accounting for the efficiency corrections discussed above, we obtain the constraints on the Model (a) parameters in Fig.~\ref{fig:1}.

We observe that since for a fixed $\cV$ mass, 
 B physics constrains the product of couplings $g_b g_\tau$, the effect in $\mathcal B_{\tau_h}$  (or equivalently in this model the modification of the total top width $\Delta \Gamma_t$) increases towards {\it smaller} values of $g_\tau$. 
 This leads to relevant constraints on the model parameter space for $m_\cV\lesssim 160$~GeV bounding $g_\tau$ from below.

\begin{figure}[!t]
\centering
\includegraphics[angle=0,width=10cm]{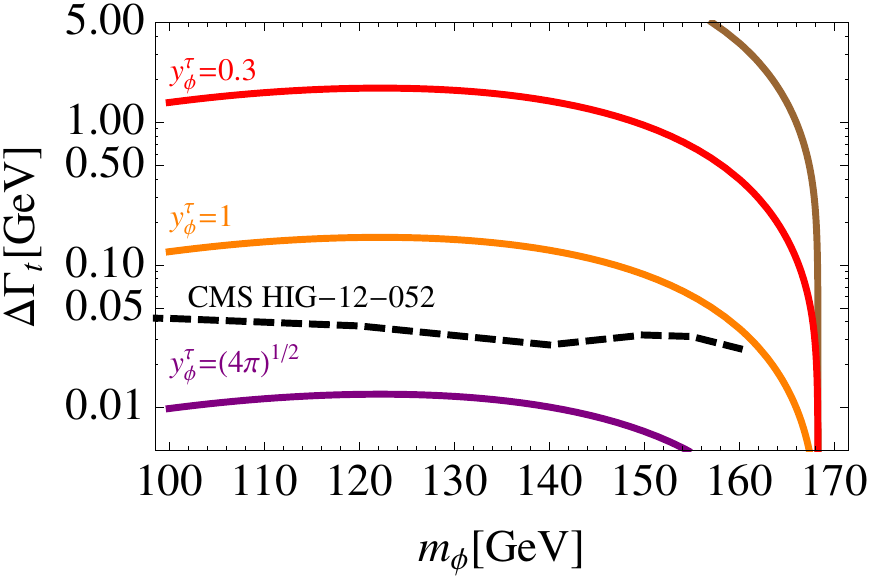}
\caption{\label{fig:2} The NP contribution to the total top width in the simplified Model (b). For each value of 
$y^\tau_\phi$ we choose the maximal possible value of  $y^{L,R}_\phi$ to satisfy the current B physics constraints. The black dashed line indicates the 
 constraints from the CMS search for the top decays to charged scalars~\cite{CMS-PAS-HIG-12-052}. See text for details.}
\end{figure}

On the other hand, dedicated searches for top decays to charged scalars in turn decaying to $\tau$ leptons ($t\to b \phi, \phi \to \tau \nu$) have been 
performed~\cite{CMS-PAS-HIG-12-052} and can easily be applied to our dynamical models of LFU violation, in particular to Model (b) when $m_\phi \lesssim m_t-m_B$. Again fixing the products of the $\phi$ couplings to SM fermions to B physics data we obtain the constraints on the Model (b) parameters in Fig.~\ref{fig:2}.
Also in this case the bounds coming from top decays are already complementary to B decays in restricting the allowed parameters space at $m_\phi\lesssim 160$~GeV to large $y_\phi^\tau$ couplings. One can also consider direct pair-production of the mediators via EW processes, with subsequent decay to $\tau \tau +{\rm MET}$. These searches exist in the context of SUSY, but due to a challenging signature and small cross sections the bounds are not 
yet competitive~\cite{CMS-PAS-SUS-17-003}.

From both Figures and also Eqs.~\eqref{eq:dB1}-\eqref{eq:dB3}, it is clear that once the NP degrees of freedom cannot be produced on-shell, current measurements of top decays become ineffective in constraining violations of LFU or respectively the related NP parameters at any appreciable level. In that regime, one can do direct searches for the NP state produced in association with third generation quarks~\cite{Aaboud:2016dig,ATLAS-CONF-2016-089,CMS-PAS-HIG-16-031,Aaboud:2018gjj,Aaboud:2018cwk}, but the limits will be model dependent. {For example, the most recent ATLAS searches for charged bosons produced in association with top and $b$-quarks, and decaying to $t\bar b$~\cite{Aaboud:2018cwk} or $\tau \nu$~\cite{Aaboud:2018gjj} are in principle sensitive to our simplified models, especially in the low mass range around 200~GeV. However, such bounds may be avoided in more complete models, if for example the on-shell mediators predominantly decay to pairs of lighter (e.g. $c\bar s$) quarks, leading to multi-jet final states.}

\section{Basic Idea: LFU probe via $b$-jet energies}
\label{sec:idea}


As we saw in the previous section, constraining LFU violating NP in top decays through 
leptonic branching ratio measurements quickly becomes infeasible. Models with NP degrees of freedom heavier than the top populate the full three-body decay phase-space while the SM predictions are dominated by two-body kinematics. This results in highly suppressed NP effects easily swamped by 
systematic uncertainties in the current LHC measurements, as well as, probably, at future colliders. 

Here we propose another strategy, exploiting precisely the kinematic 
properties of the SM top decays.  
The dominant two-body
top decays into $b$ and $W$ yield a very characteristic distribution
of $b$-quark energies in the detector frame. Neglecting for the moment the $b$-quark mass, its energy in the top rest frame is given by 
\beq\label{eq:ebstar}
E_b^* = \frac{m_t^2 - m_W^2}{2m_t}\,.
\eeq
Then, for a given boost 
$\gamma$ to the lab frame, 
if the mother particle, namely the top, is unpolarized, 
the distribution of the lab frame energies is expected to be flat between the energy 
values $E_b^* (\gamma \pm \sqrt{\gamma^2 - 1})$. This leads to a rectangular 
distribution for each given boost $\gamma$.  All the rectangles 
contain the original value $E_b^*$ which is actually
 the only energy value included in the energy distribution
for any boost.  Ref.~\cite{Agashe:2012bn} has shown explicitly that 
while the distribution itself depends on the distribution of the boosts 
$g(\gamma)$ among the 
events, the  peak of the distribution, assuming that the tops are 
unpolarized, is exactly at $E_b^*$, and that this feature 
is insensitive to the details of the function $g(\gamma)$. Since the peak of the 
distribution in Eq.~\eqref{eq:ebstar} is sensitive to the mass of the top quark, this 
allows a robust and independent determination of the 
top-quark mass~\cite{Agashe:2013eba}. Such a measurement was recently implemented by the 
CMS collaboration in Ref.~\cite{CMS-PAS-TOP-15-002}. 

The above observation is a simple consequence of the two-body kinematics and
ceases to hold for three-body 
decays (see~\cite{Agashe:2015wwa} for a detailed discussions of various 
aspects of such kinematics). In fact, in the case of the three-body decays, even 
in the rest frame of the decaying top the energy of the $b$-quark is given 
by 
\beq
E_b^* = \frac{m_t^2 - m_{l\nu }^2}{2m_t}\,,
\label{eq:Estar}
\eeq 
where $m_{l\nu}$ is the invariant mass of the lepton and the neutrino, 
which will vary across events. 
Therefore, the energy distribution of the $b$-quarks in the three body decays 
is fundamentally different from the two-body ones.   


In the case of a heavy mediator that contributes to the LFU violating top decays, the effects of the induced three-body decays (either direct or via the interference with the SM two-body decay) will manifest in small deviations from the SM in the distribution 
of the $b$-quark energies. The peak of the energy distribution will be essentially unmoved by new physics as shown in App.~\ref{app:peak}. 
On the other hand, the distribution around the peak does change more significantly due to the different kinematics of the events convoluted by the boosts that pass the kinematic cuts. This feature is less robust than the peak location, and therefore, unlike in the top mass measurement~\cite{Agashe:2013eba,CMS-PAS-TOP-15-002}, we will have to leverage some control over the boost distribution of the events. As we will later show however, it can nonetheless be highly sensitive to the presence of LFU violating charged currents in the top sector. 

\begin{figure}
	\centering 
	\includegraphics[width = .49\textwidth]{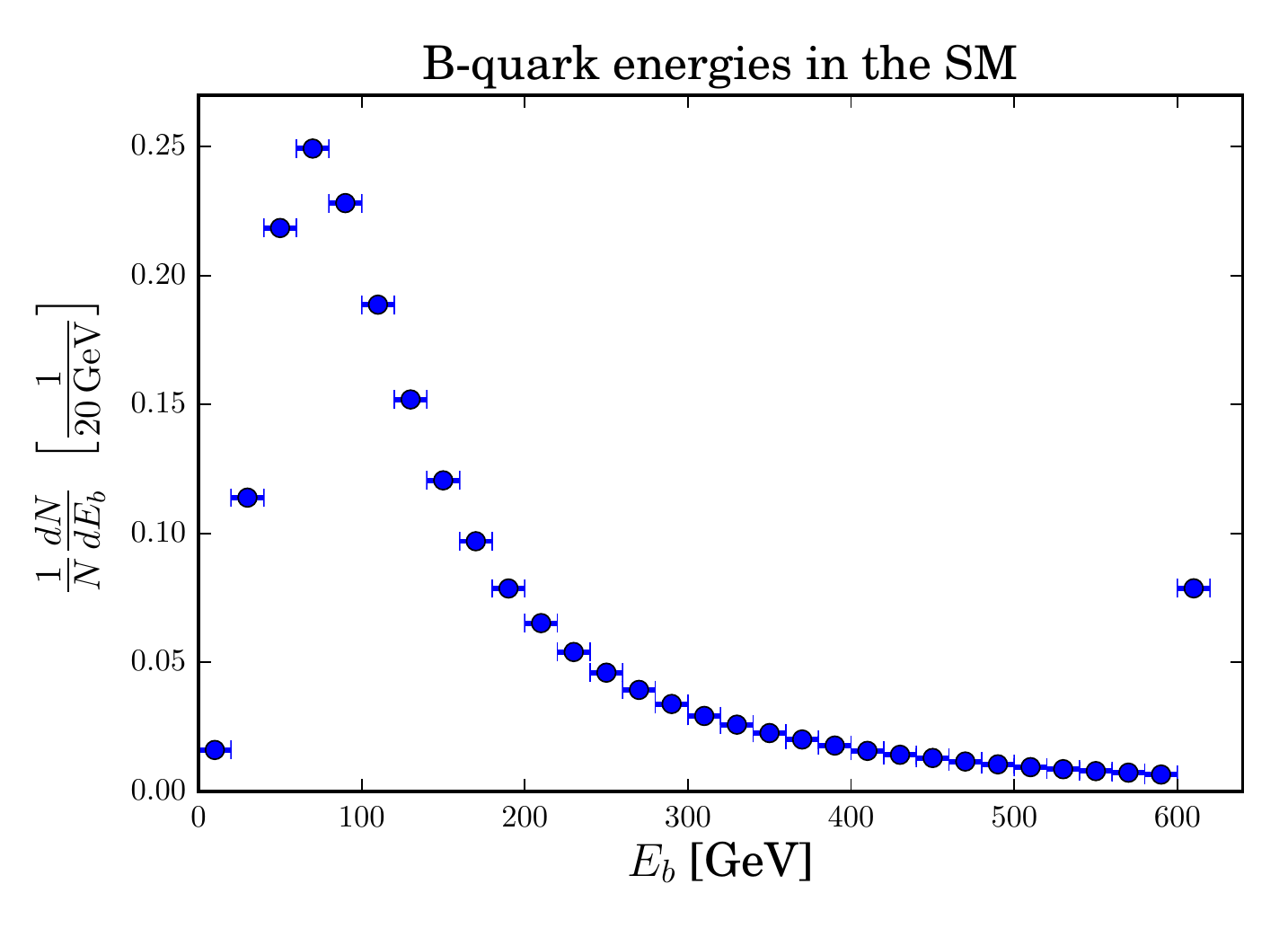}
	\includegraphics[width = .49\textwidth]{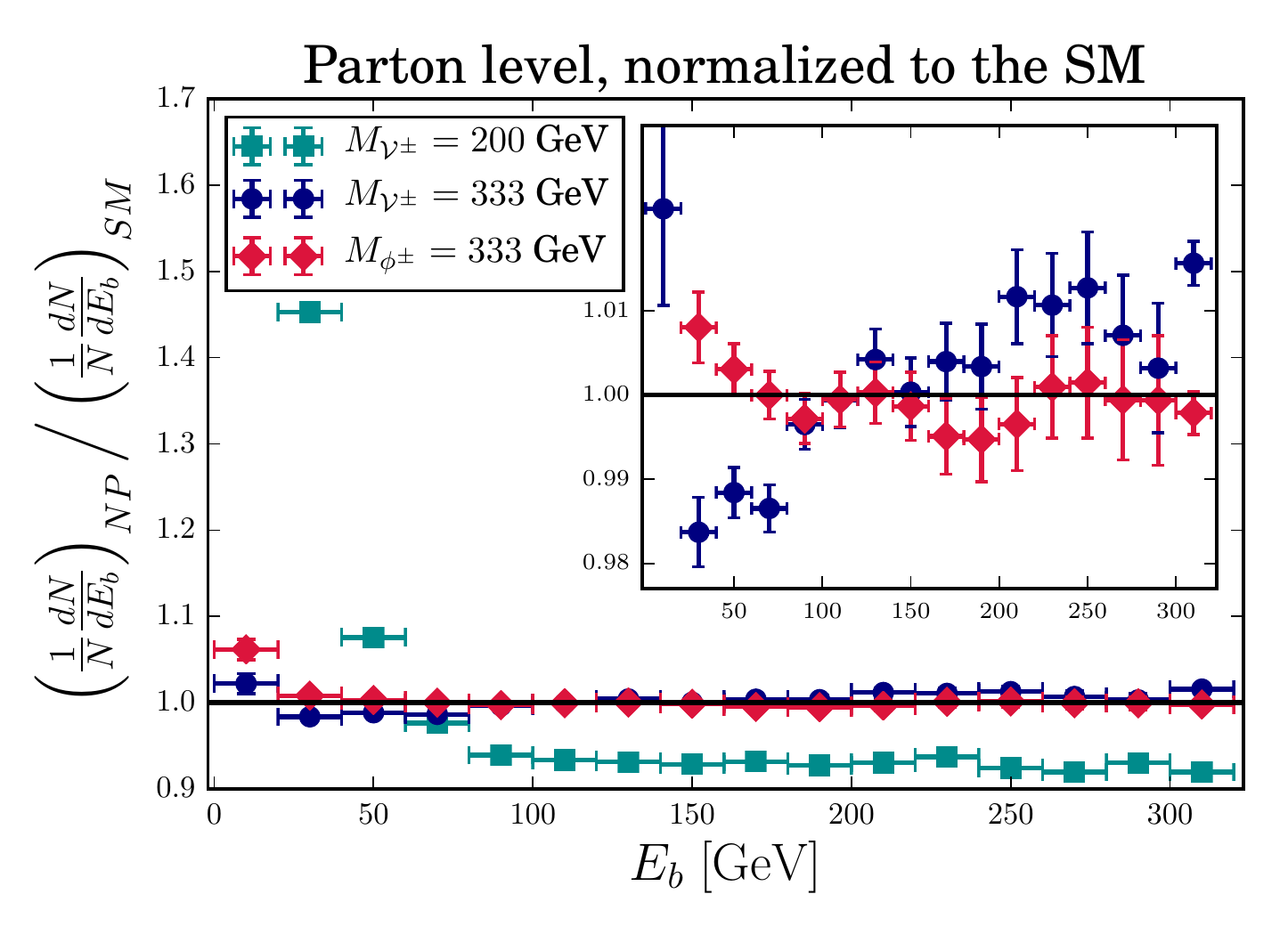}
	\caption{{\bf Left}: the normalized distribution of the lab frame $b$-quark energies in SM top quark decays simulated at parton level coming from top pair production at the $13$~TeV LHC. {\bf Right}: the ratio of normalized $b$-quark energy distribution in NP to the same distribution in the SM. We use the following NP scenarios: Model (a) with $m_{\cV}=333$~GeV and $g_\tau g_{b} = 4.5$ (blue circles), Model (a) with $m_{\cV} = 200$~GeV and $g_\tau g_{b} = 5$ (green squares), and Model (b) with $m_\phi = 333$~GeV and $y_\phi^L y_\phi^\tau = -2.6 $, $y_\phi^R y_\phi^\tau = 3.1 $ (red diamonds). The inset plot shows the data of the blue and red series magnified. All the error bars are 
statistical based on two million simulated Monte Carlo (MC) events, where Poisson statistics is assumed and with vanishing correlation between the bins. The last bin in all distributions includes overflow. }
\label{fig:parton_Eb}
\end{figure}

We demonstrate the above observations at the parton level  in Fig.~\ref{fig:parton_Eb}, where we have simulated 13 TeV LHC top pair production and decays at LO in QCD using \verb!MadGraph5_aMC@NLO!~\cite{Alwall:2014hca}. First we notice that
the peak of the $b$-quark energy distribution is around 68~GeV, as expected from Eq.~\eqref{eq:ebstar}.  We also consider our NP models (a) and (b), 
where the top is also allowed to decay via an off-shell vector or scalar boson, respectively. We plot the corresponding $b$-quark energy distributions, normalized to the SM one on the right plot for the model parameters $m_{\cV} = 200$~GeV and $g_\tau g_{b} = 5$ (in green), $m_{\cV}=333$~GeV and $g_\tau g_{b} = 4.5$ (in blue) and $m_\phi = 333$~GeV and $y_\phi^L y_\phi^\tau = -2.6 $, $y_\phi^R y_\phi^\tau = 3.1 $ (in red).
In these examples, the NP effects on the total $t\to b \tau \nu$ branching ratio are $\delta \mathcal B_{\tau} = 4\%$, 0.3\% and $0.1\%$, respectively. Except possibly for the first scenario, these effects are too small to be detected directly even at HL LHC.

As expected, in a fraction of events with NP contribution the lepton and the neutrino do not reconstruct the $W$ mass, and the $b$-quark energy distributions shift around the peak. The most affected bins are those at relatively low energies. This is generally compensated by a broad, less pronounced deficit or excess, depending on the NP model, in the higher energy bins (cf.~right panel of Fig.~\ref{fig:parton_Eb}). 
While the differences between the distributions look promising, 
as we will see in Sec.~\ref{sec:details}, discernible effects in a more realistic analysis with hadronic $b$-jets will be suppressed due to experimental acceptances and other sources of systematics that we will discuss in detail. Finally, since we only have limited theoretical control over the $b$-quark energy distributions in top pair production and decays, we will exploit the strategy of the right panel of Fig.~\ref{fig:parton_Eb} by comparing the $b$-energy distributions \emph{directly among datasets with different lepton flavors}. 

In our further analysis we assume that NP contributions to the decays of the tops into light leptons
can be safely neglected. The main idea of the analysis that we propose is then as follows: 
 tightly select all the $\tau_h \ell$ top decays, where $\tau_h$ denotes a $\tau$-tagged hadronic jet and $\ell=e,\mu$, and compare 
the resulting distribution of the $b$-jet energies to the one measured in the $e \mu$ top decay sample. The $e \mu $ channel is chosen as the cleanest one, least contaminated by non-$t\bar t$ backgrounds. 
Therefore, we do not have to cut on lepton invariant mass in the $Z$-window, which could potentially bias the $b$-jet distributions.  Essentially we look for features in the ratio between the $b$-jet energy distributions of the $\ell \tau_h$ sample and the $e\mu$ sample, similar to the right plot in Fig.~\ref{fig:parton_Eb}. 

Since we would like to tightly control the boost distribution of the $\ell \tau_h$ signal and the 
$e\mu $ control sample, we have to worry 
about systematic biases of this procedure. The most important effects come from experimental selection of events which is in general different for samples with different lepton flavors. These effects are:
\begin{itemize}
	\item One cannot reliably detect hadronic $\tau$'s with $p_T \lesssim 30$~GeV, while the threshold 
	for the detection of the light leptons 
	is typically much lower. Given small, but important, correlations between the $p_T$ of the leptons and 
	the energies of the $b$'s in $t \bar t$ events, we must make sure, that this selection bias does not propagate to the $b$-energies.     
	\item The $\ell\tau_h$ sample is expected to be  contaminated by the semileptonic $t\bar t$ events, those where one of the tops decays hadronically. The uncertainty 
	of this contamination is directly proportional to the uncertainty on the $j \to \tau_h$ mistag rate. This sample, due to very different event kinematics, 
	has a pronounced shape in the $b$-jet energy distribution. 
	\item The sample is also contaminated by a subdominant background of $(Z \to \tau^+ \tau^-)b \bar b$ events. While the cross section 
	of this background is very small compared to $t \bar t$, the resulting $b$-jet energy distribution also has a shape that differs significantly from the signal. 
\end{itemize}
In the next section we  discuss an explicit implementation of our procedure that allows to keep the above mentioned systematic uncertainties under control using data-driven methods.

\section{Implementation and discussion of uncertainties}
\label{sec:details}

\subsection{Details of simulations}
\label{sec:simulations}
Since our  strategy and main findings are based on MC simulated event samples, we here describe the simulation procedures in detail.
We have simulated our MC samples with \verb!MadGraph5_aMC@NLO!~\cite{Alwall:2011uj,Alwall:2014hca} 
at parton level and further showered and hadronized them with {\tt Pythia~8}~\cite{Sjostrand:2014zea}.

We simulate $pp \rightarrow t\bar{t} \rightarrow b\bar{b}\tau\ell 2\nu$ as our main signal processs. As we will see, we will need to properly model extra jet radiation, so we simulate samples matched up to two additional jets using the MLM-type matching~\cite{Hoche:2006ph} with a matching scale of 30~GeV. We also simulate fully leptonic $t\bar{t}$ decays in the same way. In addition, we simulate the semileptonic process: $pp \rightarrow t\bar{t} \rightarrow b\bar{b}\ell \nu 2j$, but we do not match the process since there are already extra jets in the hard process. 

We also simulate new physics contributions to top decays 
	including interference effects using the simplified models described 
	in Sec.~\ref{sec:simplified} with the model implementation described 
	in Sec.~\ref{sec:bounds}. Simulation of new physics is significantly more 
	computationally expensive than the SM because of the large number of 
	additional diagrams. Therefore, a matched NP sample is beyond our 
	technical capabilities. We get around this by approximating a NP observable 
	$O$ (such as $b$ energy) 
\begin{equation}
O^{\rm NP}_{\rm match} \approx \frac{O^{\rm SM}_{\rm match}}{O^{\rm SM}_{\rm no \;match}} \;O^{\rm NP}_{\rm no \;match},
\label{eq:match_corr}
\end{equation}
where $O$ can be any binned observable. In other words, we apply a \textit{bin by bin} correction 
	using the SM sample to account for the affects of matching. The observable will 
	then have all relevant cuts factored in for all cases.

We also simulate $pp\rightarrow Z b\bar{b} \rightarrow \tau^+\tau^- b\bar{b}$ as 
the dominant non-top background. In order to account for radiation, we match with 
one additional jet in the four flavor PDF scheme. 
We also simulate in the five flavor PDF scheme using an inclusive matching procedure. These two procedures agree in the cross section to within 20\%, and while the 
spectra are somewhat different, we get the same final results using either procedure.
 All plots are shown using the four flavor scheme.

We now detail our reconstruction algorithms. Jets are clustered with {\tt FastJet}~\cite{Cacciari:2005hq,Cacciari:2011ma}, using anti-$k_T$
algorithm~\cite{Cacciari:2008gp} with $R = 0.5$. We identify isolated leptons as those which carry away $90\%$ of the $p_T$ of all visible particles within a cone of $R=0.4$. 
We also need to identify both $b$- and hadronic $\tau$-initiated jets. 
 The experimental $b$-tagging procedures are somewhat difficult to mimic in our simulations. Furthermore, the details are less relevant, since we do not expect 
 significant signal contamination from backgrounds without $b$-jets. 
 Thus, we simply identify jets 
within $R < 0.3$ of any $b$-parton as $b$-jets.\footnote{In 
those rare cases when there is more than one jet satisfying this criterion, we 
choose the closest one to the $b$ parton. } 

On the other hand, a more realistic description of $\tau$-tagging is important for our purpose.  We thus define the following procedure to ``tag''  hadronic $\tau$'s, largely using the logic
of Ref.~\cite{Katz:2010iq}.  
First, we only consider jets with $p_T > 30$~GeV and either one or three charged tracks (prongs) within the $R < 0.08$ radius around the jet axis. We further
demand that the $p_T$ sum of all the objects within the small cone $R < 0.08$ 
around the jet axis exceeds the $p_T$
sum of all the objects within the isolation annulus of $0.08 < R < 0.4$ around 
the small cone by at least a factor of nine. This approach is of course 
still rather simplistic compared to the the experimentally used 
algorithms~\cite{Khachatryan:2015dfa,ATL-PHYS-PUB-2015-045,Flechl:2017bse}. Nonetheless, it  captures the essential features of the CMS algorithm, which is narrow jet isolation~\cite{Khachatryan:2015dfa}. Our algorithm
achieves a tagging rate of $71.3\pm 0.4\%$ with the mistag probability 
around $5.0\pm 0.1$\% estimated on $l\tau$ and semileptonic channels of the $t\bar{t}$ production,
respectively.\footnote{More precisely, in order to estimate a tagging rate, we consider 
an $l \tau$ $t\bar{t}$ sample and find the fraction of jets with $p_T > 30$~GeV, $|\eta| < 2.5$, and $\Delta R < 0.3$ from a parton level $\tau$-lepton, that are tagged as a $\tau_h$. The mistag rate is estimated based on the fraction of the 
jets that were ``identified'' as $\tau_h$ in the semileptonic  $t\bar{t}$ sample as a fraction of 
all the non-$b$-jets with the same kinematic acceptance criteria as before. The uncertainty on the (mis)tag rates are due exclusively to MC statistics.}
As we further dissect these numbers, we find that in the semileptonic 
sample we have similar numbers of 1-prong and 3-prong fake 
hadronic $\tau$'s.  Among the tagged $\tau$'s in the $l\tau$ sample, we find that 1/3 of all the 
hadronic $\tau$'s are 3-prong, which is comparable to the true branching ratio of hadronic $\tau$'s.  
The mistag rates in our simulation are significantly higher than current state of the art experimental taggers. Therefore, one can think of our $\tau$-tagging procedure as extremely conservative, 
in the sense that the experimental collaborations are expected to perform better than our 
simulations. 

As we will further see, one of the most important backgrounds in our analysis is the semileptonic 
$t \bar t$ where one of the jets is misidentified as a $\tau$-jet. Even though our $\tau$-tagging is very conservative, 
this is likely to be an important background also in a realistic analysis. For example, in recent experimental analyses of tauonic top 
decays~\cite{Khachatryan:2014loa,Aad:2015tin,Aad:2015dya},  
the semileptonic $t \bar t$ was identified as the dominant background. We reduce the amount of  non-$\tau \ell$ events in the signal sample due to this background by vetoing extra jets in the final state. For this purpose we match all our leptonic $t \bar t$ SM samples to parton shower with up to two additional jets. We do not know whether the jet veto will be necessary in a realistic search, where the hadronic tau mistag rate is much smaller than what we get, nonetheless in this search we perform it in order to demonstrate the viability of our procedure even with extremely unfavorable assumptions.

\subsection{Analysis}
\label{subsec:analysis}

We begin by imposing the following selection criteria for the signal events:
\begin{itemize}
	\item Exactly one isolated light lepton ($\ell  = \mu$ or $e$) with $p_T > 20$~GeV and $|\eta| < 2.5$\,.
	\item Exactly one $\tau$-tagged jet with $p_T > 30$~GeV and $|\eta| < 2.5$\,.
	\item Exactly two $b$-jets with $p_T > 20$~GeV and $|\eta|< 2.5$\,.
\end{itemize}
The efficiency times acceptance of this selection is around 7\% for our signal sample ($\ell \tau$ signature decays of $t \bar t$). 
For future reference we define 
\begin{equation}
n[{\rm signature}](E_b) \equiv \frac{N[{\rm signature}](E_b)}{\sum_{(E_b)} N[{\rm signature}](E_b)}\,,
\label{eq:distribution}
\end{equation}
as the relative number of $b$-jets with $b$-jet energies within the bin $(E_b)$ in events with a given experimental [signature]. Note that each event passing our selection cuts contributes two $b$-jets to the sample. In the case of the signal, the relevant quantity is thus $n[\ell \tau_h 2 j_b]$.

We now need to compare our signal to a control sample where the contribution from new physics is suppressed. We use dileptonic $t\bar{t}$ decays with opposite flavor leptons. Namely, we require one isolated electron, one isolated muon, and zero $\tau$ with the same kinematic requirements as above.  Such a selection introduces an immediate bias that can swamp potential NP effects  we are looking for. The problem 
is that the energy and momentum of a hadronically decaying $\tau$ are shared among the resulting $\tau$-jet and the (undetected) $\tau$-neutrino. Thus, compared to a light lepton, a selected $\tau$-jet corresponds of a given $p_T$ typically corresponds to a $\tau$-lepton of much higher $p_T$. 
 Because of a non-vanishing correlation between $p_T(\tau)$ and  $E_b$, equivalent cuts on the $\tau$ $p_T$ in the signal sample and the lepton $p_T$ in the control sample will lead to different distributions of $b$-jet energies.
 
 To compensate for this effect we propose the following 
scheme: we take the selected $e\mu $ events and substitute one randomly chosen reconstructed lepton with a $\tau$ and let it  decay using MC simulation.\footnote{Note that it is important to use the correct polarization of the simulated $\tau$ because this has significant imprints on the energies of the $\tau$ decay products. In {\tt Pythia~8}, this corresponds to using \;\;\;\;\; \;\;\;\;\; \;\;\;\;\; polarization = - charge. }
Since $\tau$ decays have been measured experimentally to the level of 
much better than one percent, this should not introduce an insurmountable systematic problem.  After performing 
this substitution we apply to this ``corrected'' control sample exactly the same selection criteria as to the original $\ell \tau$
sample. In this way we obtain the control sample $n[\ell_h \ell' 2 j_b](E_b)$, where $\ell_h$ refers to a light lepton replaced in simulation by a hadronically decaying $\tau$. The efficiency times acceptance of this procedure is around 7\% based on our simulations.  

For illustration we show the ratio of the $b$ energy between the $\tau\ell$ sample and the $e\mu$ sample, namely $n[\ell \tau_h 2 j_b]/n[\ell_h \ell' 2 j_b] (E_b)$. The blue circles use the naive ratio without doing the replacement of leptons with $\tau$ in MC, and we see that the SM has the same shape as the new physics shown in Fig.~\ref{fig:parton_Eb}. On the other hand, once we apply our our replacement procedure we get the red diamonds which have very good agreement with a flat shape, signalling no appreciable differences between the signal and the control. The plots were produced using $\sim 6.3$ (6.6) million MC events for the signal (control) channels corresponding to an effective LHC luminosity of $\mathcal L \simeq 75$ (200) fb$^{-1}$, and resulting in an  
uncertainty due to MC statistics of 1\% to 2\% per $20$\,GeV energy bin.  

\begin{figure}
	\centering
	\includegraphics[width =.49\textwidth]{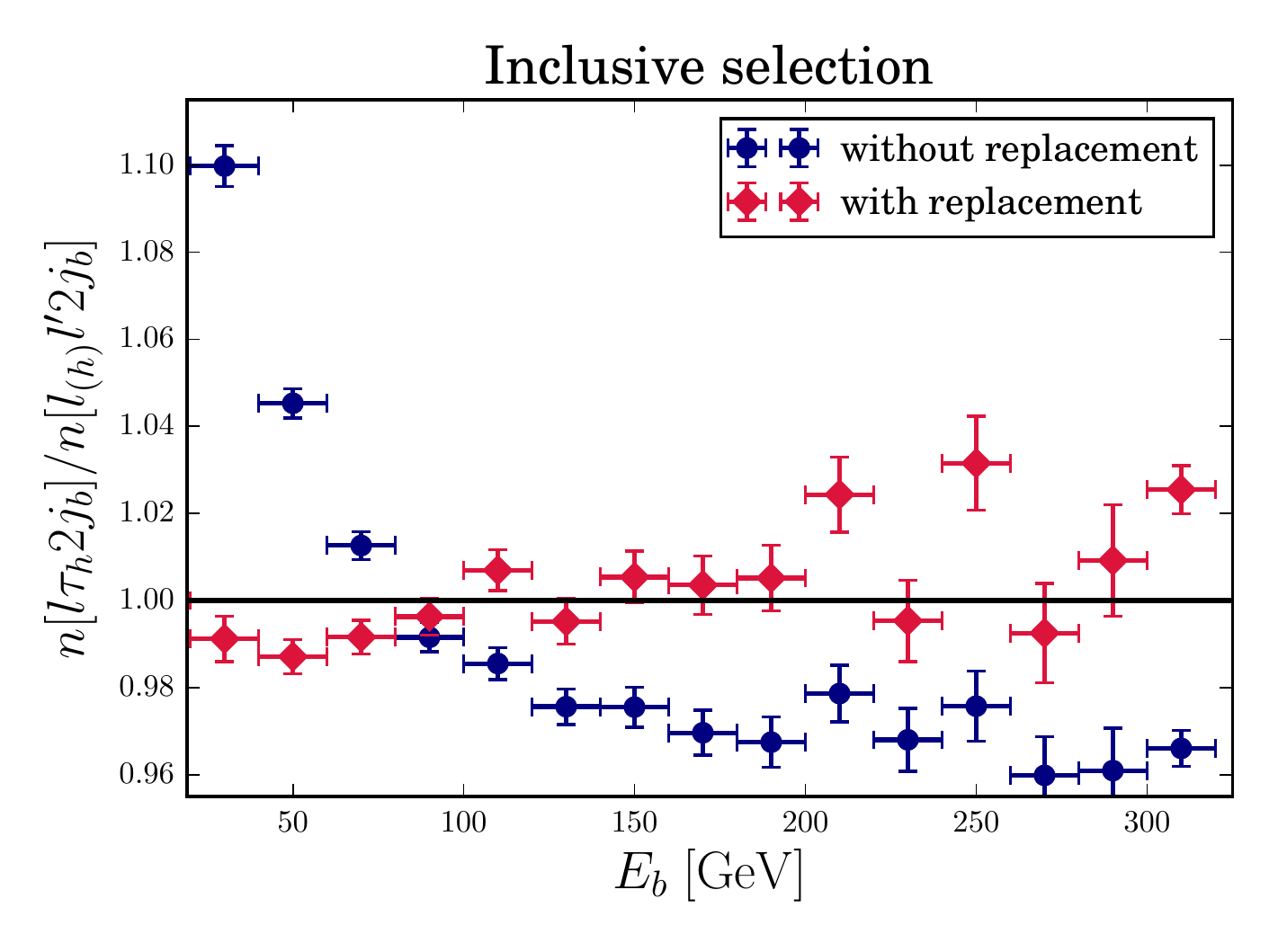}
	\includegraphics[width=.49\textwidth]{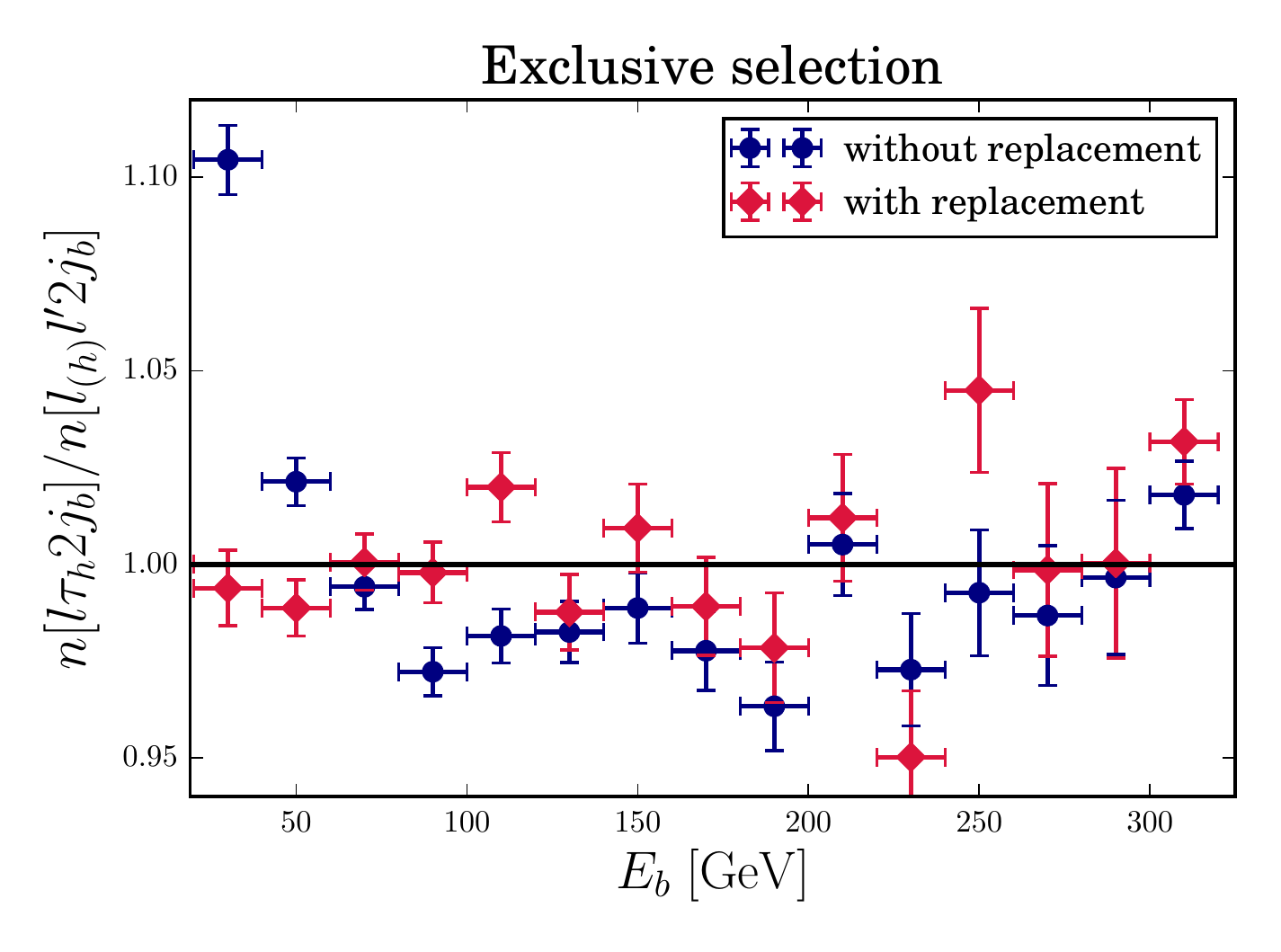}
	\caption{{\bf Left:} the distribution of the $b$-jet energies in the SM $\tau \ell$ sample, normalized to the 
		control ($e\mu$) sample as explained in the text. The blue points use a naive $e\mu$ control sample, while the red ones use the control sample with one of the leptons replaced by a $\tau$ in simulation. 
		The event selection is consistent with the baseline cuts outlined at the beginning of 
		Sec.~\ref{subsec:analysis}. {\bf Right:} the same, but with an additional jet veto and a restriction to one-prong $\tau$'s only. See text for details.}
	\label{fig:SMselection}
\end{figure}

There will also be $\tau\ell$ events that leak into the control sample. Since in roughly $1/3$ of $\tau$'s decays leptonically, some portion 
of these events will unavoidably look like $\mu e$ events. These events might be affected by the NP, so this is 
effectively a spill out of the NP into the control sample. Fortunately, due to very low acceptances of the leptonic $\tau$'s 
(due to very low momentum), the effect in not particularly big. We estimate this spill out to be 10\%  of the 
genuine $e \mu $  sample. We do not include this effect on Fig.~\ref{fig:SMselection}, but we will include it in the 
final plots.

The dominant background in the signal region will be $t\bar{t}$ with one top decaying leptonically, and the other to $bjj$ where one of the jets fakes a $\tau_h$. Because the $p_T$ of the `fake' $\tau$-jets does not correspond to the $p_T$ of the real $\tau$-jets in the signal, this is expected to introduce a non-trivial shape in the $E_b$ distribution. 
With our $\tau$-tagging procedure, the fake acceptance of these events without further cuts is 1.6\%, as compared to the 6\% for the signal. Because this background also has a larger cross section than the signal, more selection cuts are needed to mitigate this background.

\begin{figure}
	\centering
	\includegraphics[width =.7\textwidth]{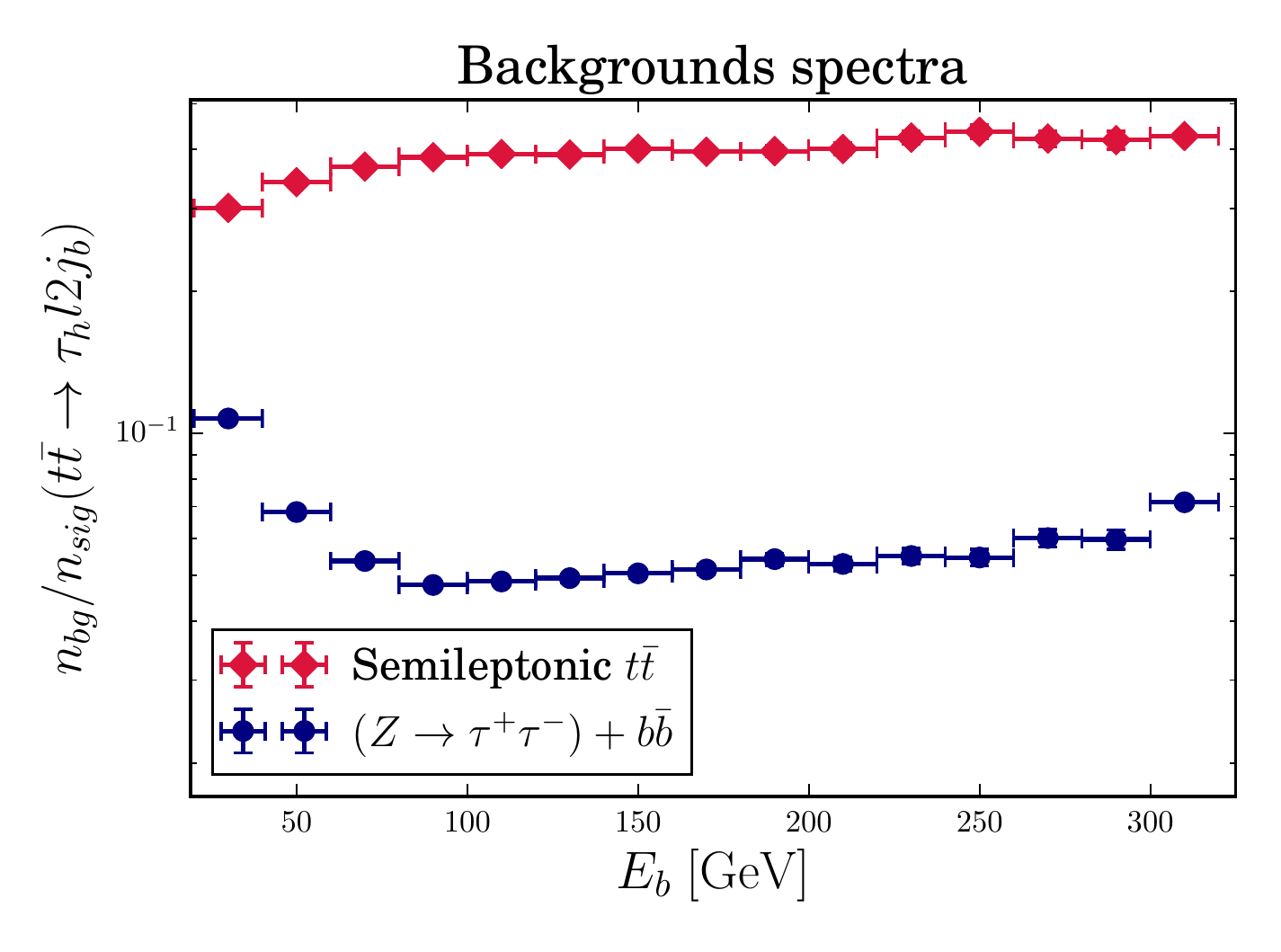}
	\caption{The distribution (on a log scale) of the $b$-jet energies in the background processes normalized to the distribution of the signal process, $t\bar{t} \rightarrow b\bar{b} \tau_h \ell 2\nu$. The red higher curve corresponds to semi-leptonic top decays, while the lower blue curve is $Z\bar{b}b$. Cross section times efficiency with the full set of cuts gives the normalization of the two background curves.}
	\label{fig:backgrounds}
\end{figure}

We address the problem in several steps. First, we impose a jet-veto on our signal and control 
samples, allowing in each event exactly 2 $b$-jets, and no non-tagged jets with $p_T > 20$~GeV.\footnote{This cut is why all leptonic samples must be generated matched to extra jets, see Sec.~\ref{sec:simulations} for more details.}  This cut is dangerous in our 
context, because it potentially biases the boost of the accepted events. However, if the jet veto is applied to the 
signal and control samples, the bias largely cancels out. Second, we restrict our analysis to
only one-prong hadronic $\tau$-jets, because that captures the majority of signal events while eliminating about half the background. After the jet veto and considering only one-prong $\tau$'s, the semileptonic $t\bar t$ background acceptance falls to $0.05\%$ while the signal and control channel acceptances are also somewhat reduced to $1.9\%$ and $2.1\%$, respectively. 
In Fig.~\ref{fig:backgrounds} we plot the $E_b$ distribution of the semileptonic background in the red diamonds with the normalization set by the cross section times efficiency of the full set of cuts relative to that of the signal. We see that even with these cuts, the semileptonic background is about 1/3 as big as the signal and it has a very different shape.

\begin{figure}[t]
	\centering
	\includegraphics[width =.49\textwidth]{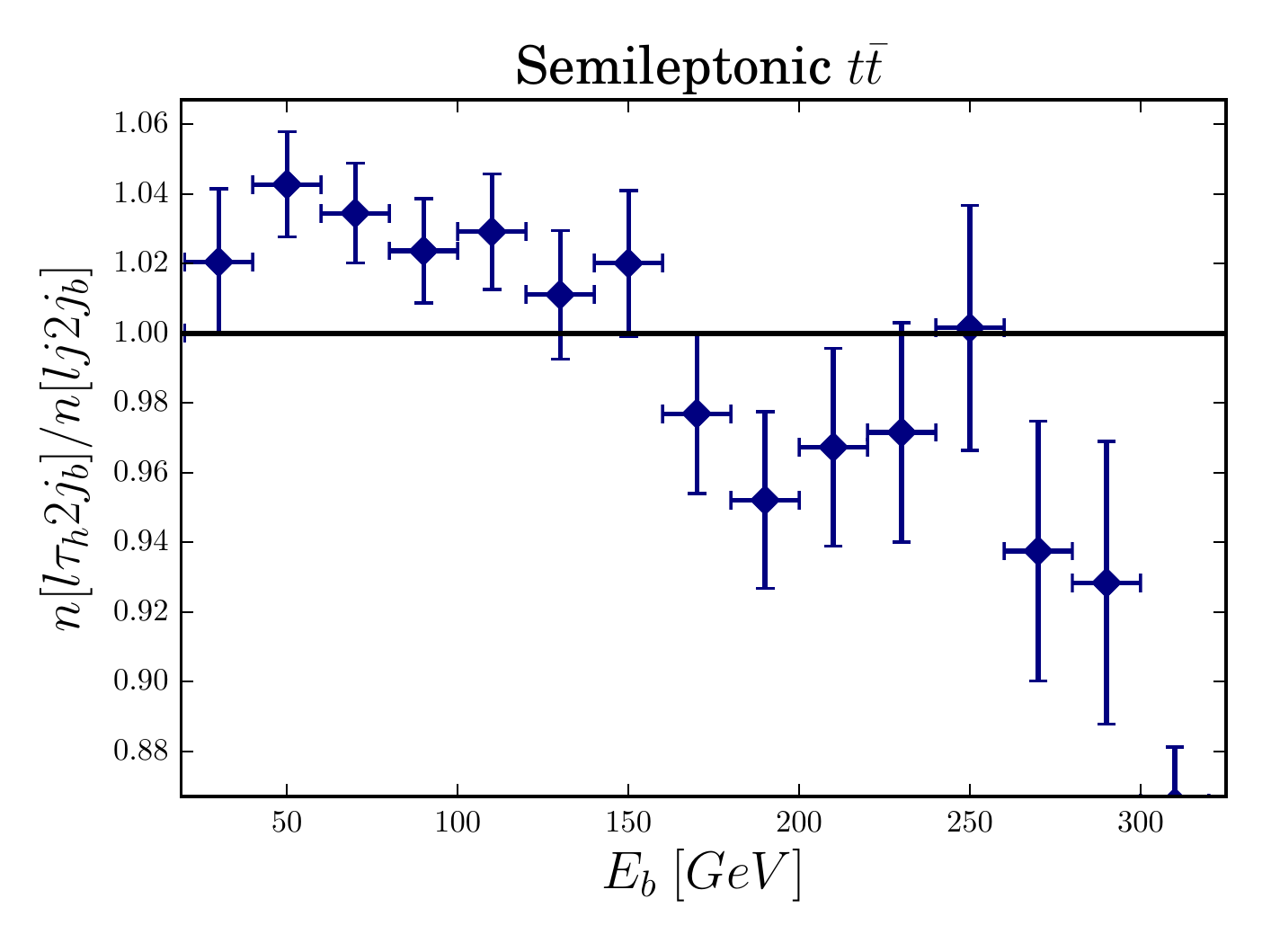}
	\includegraphics[width=.49\textwidth]{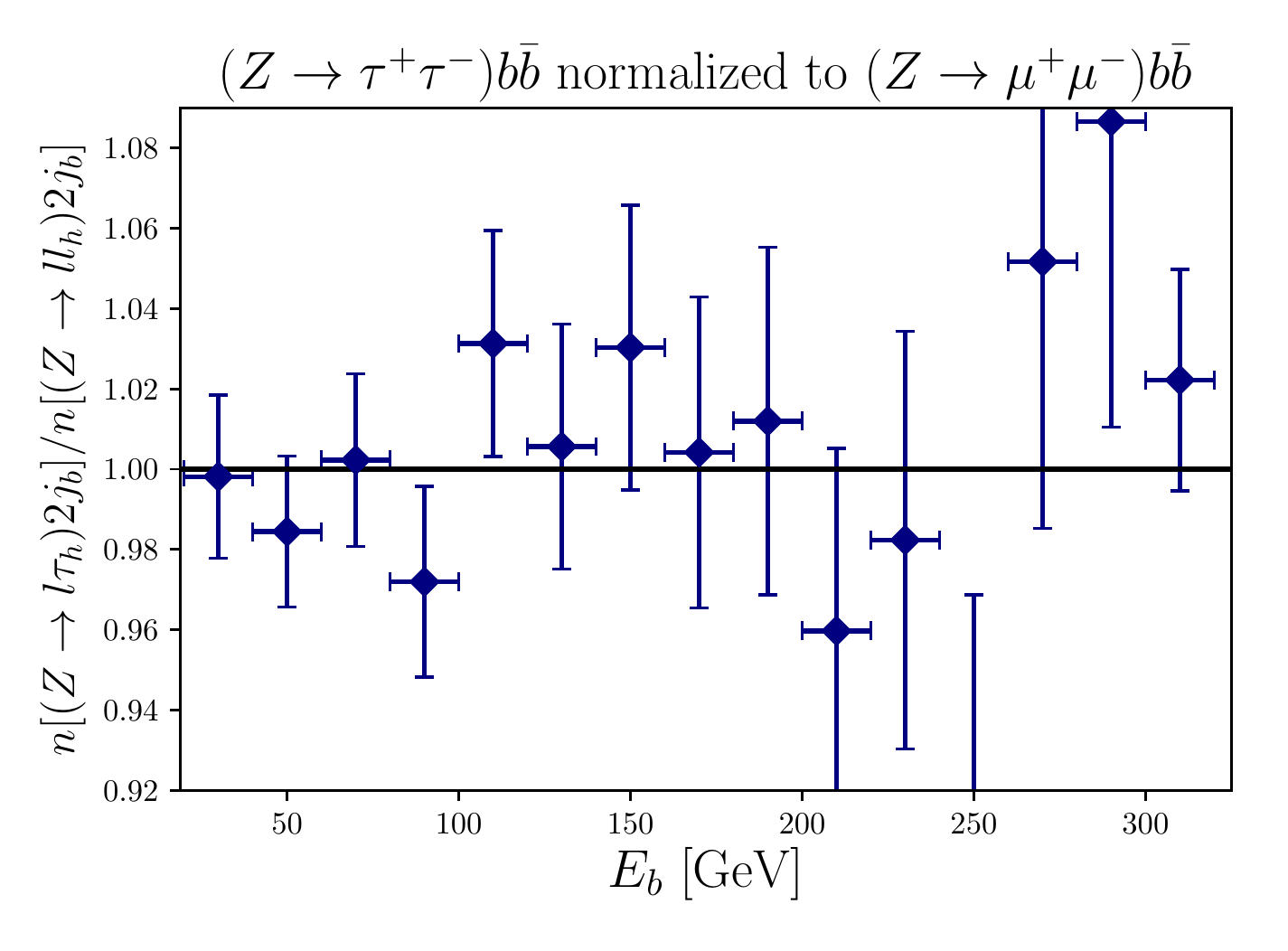}
	\caption{{\bf Left:} the distribution of the $b$-jet energies in the SM semileptonic $t\bar t$ sample where one of the jets is $\tau$-mistagged, normalized to the 
		same sample without the $\tau$-tag. 
		{\bf Right:} the distribution of the $b$-jet energies in the SM $(Z \to \tau \tau)b \bar b$  sample where one of the taus decays muonically and the other is $\tau$-tagged, normalized to the 
		$(Z \to \bar \mu \mu) b \bar b$ sample, where both of the muons are replaced by a $\tau$ in simulation. See text for details.}
	\label{fig:background_ratio}
\end{figure}

To account for this shape, we propose one more trick: to add a second control sample to our original one. The second control sample is events with 
\begin{itemize}
\item exactly one lepton, zero $\tau$ tagged jets,
\item two $b$-tagged jets, 
\item one non-$b$-tagged jet with $p_T > 30$ GeV,
\item zero additional jets with $p_T > 20$ GeV.
\end{itemize}
As we are demanding that our $\tau$-jets be one-prong, three-prong $\tau$'s are counted as ordinary jets. 
As a cross-check of this procedure, we show the $b$-jet energy distribution of the ratio $ n[\ell (j \to \tau_h) 2 j_b]/ n[\ell j 2 j_b](E_b)$ on the left side of Fig.~\ref{fig:background_ratio}. 
 While the additional control sample clearly improves the situation, it unfortunately
 does not fully get rid of the shape. We are left with a manageable systematic uncertainty in the most relevant
low energy bins, and a larger uncertainty in the high energy bins. We hope that experimentalists will find a more refined solution to better account for this background 

We combine the two control samples in such a way that the fraction of semileptonic events in 
the $\tau \ell$ signal sample, $w_{j \to \tau_h} \simeq 0.4$, is the same in the control sample. Namely, our 
control sample is $n[\ell'_h \ell 2 j_b](E_b) + w_{j \to \tau_h} n[\ell j 2 j_b](E_b) $, 
and $w_{j \to \tau_h}$  includes the ratio of the relevant decay branching fractions and $\tau$ (mis)tag rates.

The other important background is $(Z \to \tau^+ \tau^+) + b \bar b$ where one $\tau$ decays leptonically and the other hadronically. It is very small, having a cross-section 
of $\sim 15$~fb after our cuts. Because the $b$-jets are a result of radiation, their energies are peaked towards the lowest values allowed by experimental cuts. Therefore, this background has a highly non-trivial shape which can be seen as the blue circles in 
Fig.~\ref{fig:backgrounds}. To account for this process, we add a third control region,  $ n[(Z \to \ell_h \ell_h)  2 j_b](E_b)$, namely 
we simulate leptonic decay of the $Z$ and then replace both leptons with $\tau$'s in simulation. This control sample 
has corresponding weight $w_{Z \to \ell \tau_h} \simeq 0.03$ which takes into account the ratio of the 
$Z b \bar b$ and $t\bar t$ cross-sections passing our cuts and also leptonic $Z$ branching fractions. We note that in the 
ratio the relatively large QCD K-factors to the simulated production cross-sections
 $K_{\rm NNLO}(t\bar t) = 1.8$~\cite{Czakon:2013goa,Czakon:2011xx} and 
 $K_{\rm NLO}(Z b\bar b) = 1.7$~\cite{Frederix:2011qg,Alwall:2014hca} almost cancel. Again for cross-check, we 
 show the relevant ratio of $b$-jet energy distributions for this background $ n[(Z \to \ell \tau_h) 2 j_b]/ n[(Z \to \ell \ell_h) 2 j_b](E_b)$ on the right side of Fig.~\ref{fig:background_ratio} and there appears to be no discernible shape within the statistical MC uncertainties. This process also contributes to the $e\mu$ control sample when both $\tau$'s decay leptonically, but this contribution is sub per mille because the branching ratios are reduced and the leptons tend to be quite soft.
Our signal and control regions, processes that contribute to each, and the relevant cross sections and efficiencies are summarized in Tab.~\ref{tab:efficiencies}.

\begin{table}[tb]
\centering
\begin{tabular}{|c|c|c|c|c|}
\multicolumn{5}{c}{ Signal [$\ell \tau_h 2 j_b$] } \\ 
\hline
Process & $N_{\rm MC}$ & $\sigma$ (pb) & $\epsilon_{\rm inc}$(\%) & $\epsilon_{\rm ex}$(\%) \\ \hline
$t\bar{t} \rightarrow b\bar{b} \tau \ell 2\nu$ & 6.3M  & 84.2 & 6.7  & 1.9   \\ \hline
$t\bar{t} \rightarrow b\bar{b} \ell \nu 2j$ & 40M  & 416  &  1.6 & 0.046 \\ \hline
$Z(\rightarrow \tau \tau) b\bar{b}$ & 5.4M  &  4.79 &  1.2 &  0.32 \\ \hline
\end{tabular}
\qquad\qquad
\begin{tabular}{|c|c|c|c|c|}
\hline
Process & $N_{\rm MC}$  & $\epsilon_{\rm inc}$(\%) & $\epsilon_{\rm ex}$(\%) & $w$\\ \hline\hline
\multicolumn{5}{|c|}{ CR [$\ell_h \ell' 2 j_b$] } \\ \hline
$t\bar{t} \rightarrow b\bar{b} \ell \ell' 2\nu$ & 6.6M  &  7.4   & 2.1 & 0.908 \\ \hline
$t\bar{t} \rightarrow b\bar{b} \ell \tau 2\nu$ &  7.6M  & 0.33  & 0.087 & 0.092 \\ \hline\hline
\multicolumn{5}{|c|}{ CR [$\ell j 2 j_b$] } \\ \hline
$t\bar{t} \rightarrow b\bar{b} \ell \nu 2j$ &  40M  & 28  & 4.2 & 0.42  \\ \hline\hline
\multicolumn{5}{|c|}{ CR [$Z(\rightarrow \ell_h \ell_h) 2 j_b$] } \\ \hline
$Z(\rightarrow \ell \ell) b\bar{b}$ & 5M &  1.2   & 0.32  & 0.033 \\ \hline
\end{tabular}
\caption{The production cross section, number of generated MC events, and the acceptance rates 
(in the inclusive and the exclusive samples respectively) of our signal process and the background processes. On the 
left hand side we show the signal and the two dominant backgrounds, namely the semilepronic $t \bar t$ and 
$(Z \to \tau^+ \tau^-) b \bar b$. On the right hand side we show the control regions with the appropriate weights $w_i$
as they are defined in Eq.~(\ref{eq:master}). }
\label{tab:efficiencies}
\end{table}

\subsection{Results }

\begin{figure}
	\centering
	\includegraphics[width=.7\textwidth]{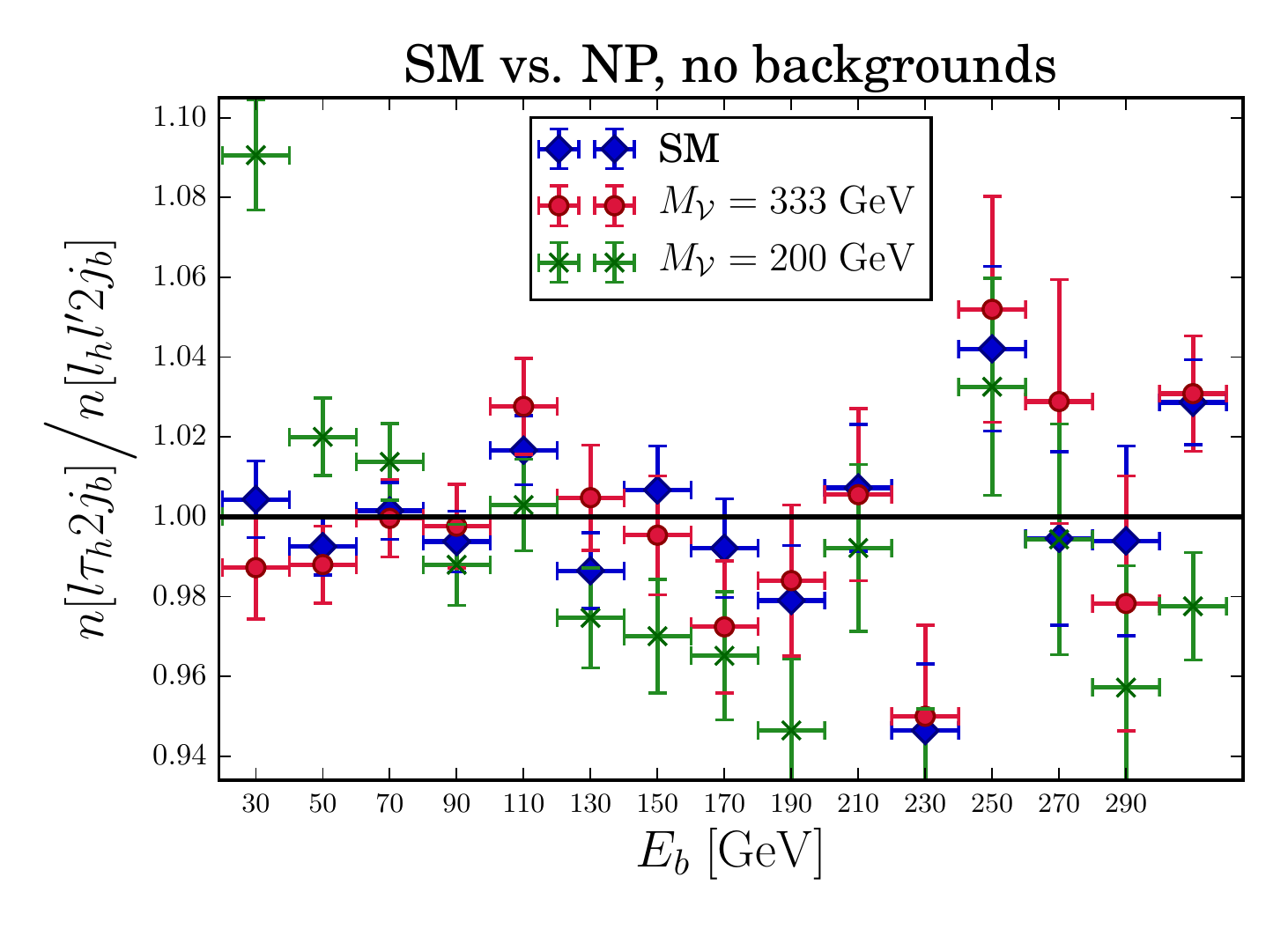}
	\caption{ The $b$-jet energy binned distributions of the lepton universality ratio $n[\ell \tau_h 2 j_b]/n[\ell'_h \ell 2 j_b]$ defined in Eq.~(\ref{eq:distribution}), without any background included, in the SM (in blue) as well as in the simplified NP Model (a) with $m_{\cV}=333$~GeV and $g_\tau g_{b} = 4.5$ (in red), and with $m_{\cV} = 200$~GeV and $g_\tau g_{b} = 5$ (in green). See text for details.}
	\label{fig:pre_money}
\end{figure}

\begin{figure}
	\centering
	\includegraphics[width=.7\textwidth]{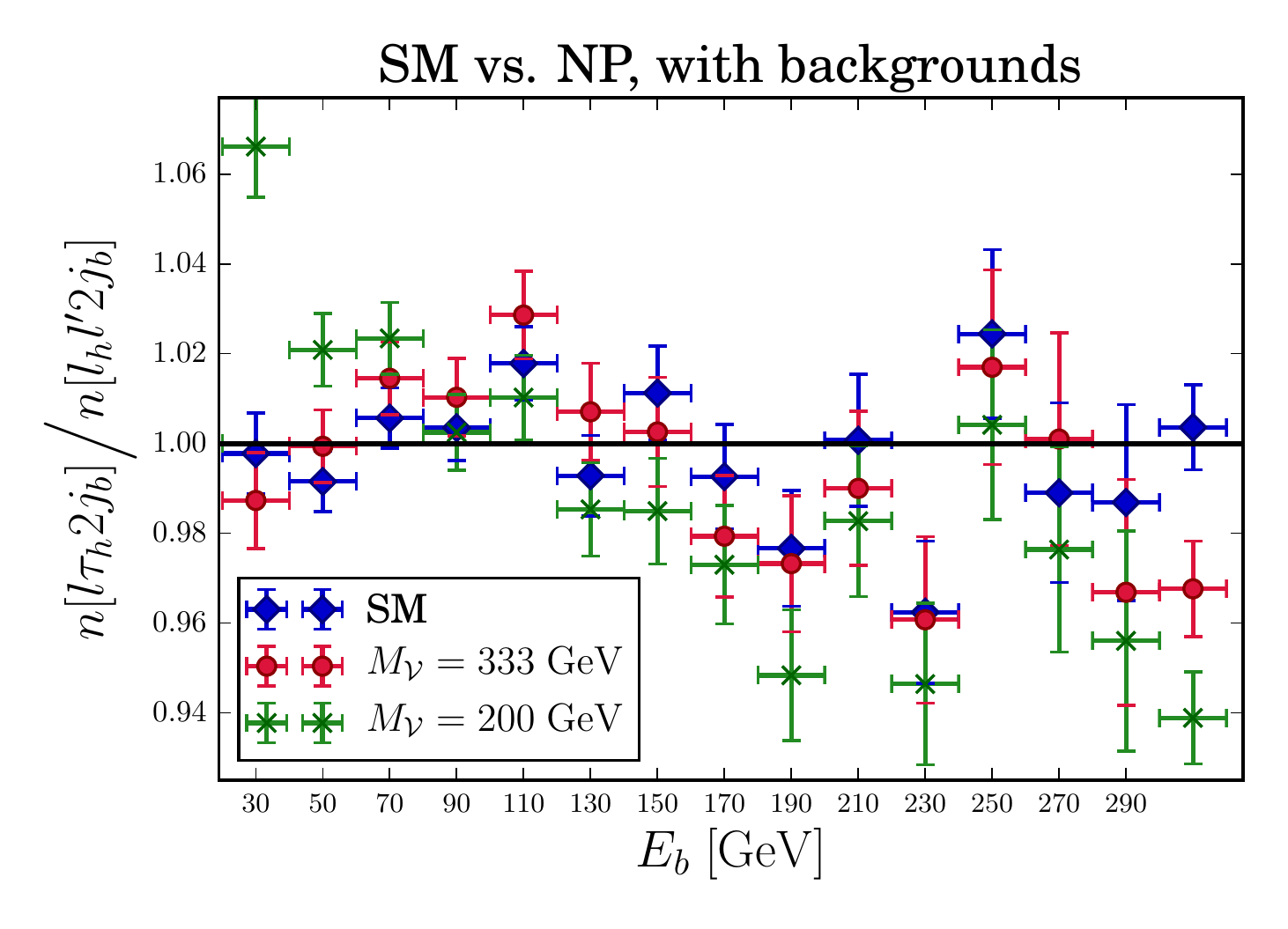}
	\caption{ The $b$-jet energy binned distributions of the lepton universality ratio $R_{\tau_h/\ell}(E_B)$ defined in Eq.~\eqref{eq:master} in the SM (in blue), as well as in the simplified NP Model (a) with $m_{\cV}=333$~GeV and $g_\tau g_{b} = 4.5$ (in red), and with $m_{\cV} = 200$~GeV and $g_\tau g_{b} = 5$ (in green). In this figure we add
		all relevant backgrounds with the proper subtraction procedure as explained in the text. The differences 
		from Fig.~\ref{fig:pre_money} are, as expected, not significant, since both backgrounds have been tamed by 
		appropriate control regions. }
	\label{fig:money}
\end{figure}

In Figs.~\ref{fig:pre_money} and~\ref{fig:money}, we compare the SM predictions (in blue) to our NP benchmark models:
\begin{itemize}
\item $m_{\cV}=333$~GeV and $g_\tau g_{b} = 4.5$ (in red),
\item $m_{\cV} = 200$~GeV and $g_\tau g_{b} = 5$ (in green). 
\end{itemize}
In Fig.~\ref{fig:pre_money} we plot $n[\ell \tau_h 2 j_b]/n[\ell'_h \ell 2 j_b]$ using the full event selection for the three samples with no backgrounds included. We see that within the errors, the SM is consistent with one across the distribution, and the scatter around the flat distribution should be viewed as a measure of our systematic uncertainties due to limited 
MC statistics and not having a sufficiently accurate control sample for our semileptonic background. Unfortunately, it appears that the first benchmark NP scenario with $m_{\cV}=333$ GeV is also consistent with one. Our other benchmark with  $m_{\cV} = 200$~GeV, however, shows the 
characteristic steep rise at low energy and broad deficit at higher energies consistent with the parton level simulation shown on the right panel of Fig.~\ref{fig:parton_Eb}. 

\begin{table}[tb]
\centering
\begin{tabular}{|c|c|c|c|c|}
\multicolumn{5}{c}{ NP vs. SM } \\ 
\hline
Model & $\epsilon_{\rm inc}$(\%) & $\epsilon_{\rm ex}$(\%) & $\chi^2$ & $\chi^2_3$\\ \hline
SM (unmatched) & 6.82  & 1.71 &  41.3 & 4.1   \\ \hline
$m_{\cV}= 333$ GeV &  6.75 & 1.69  &  41.0 & 4.1 \\ \hline
$m_{\cV}= 200$ GeV &  7.69 & 1.93  &  147 &  61.6 \\ \hline
\end{tabular}
\caption{Efficiencies and $\chi^2$ values for various NP models and the SM. The efficiencies are for unmatched samples, and the $\chi^2$ distributions use the data shown in Fig.~\ref{fig:money}, with $\chi^2_3$ using only the first three bins. }
\label{tab:BSM}
\end{table}

We now construct our final observable taking into account all sources of background. The $b$-jet energy binned lepton universality ratio:
\begin{equation}\label{eq:master}
R_{\tau_h/\ell}(E_B) \equiv \frac{n[\ell \tau_h 2 j_b]}{n[\ell'_h \ell 2 j_b](E_b) + w_{j \to \tau_h} n[\ell j 2 j_b](E_b) 
	+  w_{Z \to \ell \tau_h} n[(Z \to \ell \ell_h)  2 j_b](E_b)}\,.
\end{equation}
We plot this variable both for the SM and the benchmark NP scenarios on Fig~\ref{fig:money}. Because the NP samples have slightly different efficiencies than the SM sample, the values of $w$ for the different control samples will change slightly, and this is taken into account in Fig~\ref{fig:money} with the different efficiencies shown in Tab.~\ref{tab:BSM}.

There are no striking differences between Figs.~\ref{fig:pre_money} and~\ref{fig:money} indicating that the background control procedure, though not ideal, works reasonably well.  Again, we see that the higher mass benchmark looks relatively similar to the SM, and the lower mass benchmark is very clearly distinguishable from a flat shape. On the other hand, the errors are dominated by Monte Carlo statistics corresponding to a data set $\mathcal{O}(100)$ fb$^{-1}$. We therefore expect with the full HL LHC dataset, the errors should shrink significantly and the reach will improve.

We also perform the following simple statistical tests on the data shown in Fig.~\ref{fig:money}. Calculating a $\chi^2$ fit to a flat distribution $R_{\tau_h/\ell}(E_B) = 1$, we obtain $\chi^2_{\rm SM} = 41.3$, which corresponds to a $p$-value with 14\footnote{There are 15 bins, but it is a normalized distribution, so it has 14 degrees of freedom. On the other hand, $\chi_3^2$ does not have the normalization condition, so it has three bins and three degrees of freedom.} degrees of freedom of $1.7\times10^{-4}$. This indicates that our errors from MC statistics alone do not fully account for our systematic error budget. A more detailed estimate of systematics would take into account other major sources, in particular the systematics arising from imperfect control samples, especially for the semileptonic background. From the left panel of Fig.~\ref{fig:background_ratio}, we see that the systematics are largest in the high energy bins, and from the right panel of Fig.~\ref{fig:parton_Eb}, we see that the effects of new physics should be largest in the low energy bins. Therefore, we also compute the $\chi^2$ distribution for only the first three bins, $\chi^2_3 = 4.1$ for the SM, implying a more sensible $p$-value with three degrees of freedom of 0.26. 

We can now compare the $\chi^2$ values for the NP models. For our benchmark with relatively smaller effects and $m_{\cV} = 333$ GeV, we get $\chi^2 = 41.0$ ($\chi^2_3 = 4.1$), nearly identical values to the SM, confirming that with this amount of data, there is no sensitivity to this benchmark. On the other hand, for our NP sample with $m_{\cV} = 200$ GeV, we get get  $\chi^2 = 147$ ($\chi^2_3 = 61.6$). Using the $\chi^2_3$ value we can compute a naive $p$-value that corresponds to approximately $7\sigma$ exclusion of that model. Alternatively, we can compute the $p$-value from $\Delta\chi^2 \equiv \chi^2_{\rm NP} - \chi^2_{\rm SM}$ which follows a variance-gamma distribution, and this $p$-value gives an approximately $9\sigma$ exclusion. The $\chi^2$ values are summarized in Tab.~\ref{tab:BSM}. While a full statistical analysis taking into account all sources of systematic error is beyond the scope of this work, these simple tests show that there is clearly sensitivity to new physics with mediator masses above the top mass.

\section{Summary and Conclusions}
\label{sec:conclusions}

In this paper we have made a case for experimental probes of possible LFU violating NP in top decays. First we have shown how LFU violation in top decays can be related to recent intriguing results in semi-tauonic B decays. The correspondence is subject to two important effects: (1) the scaling of flavor effects between $b\to c$ and $t\to b$ transitions necessarily needs to assume some flavor structure of NP. While in our analysis we have relied on the most conservative MFV-like scaling, other possibilities predicting smaller or larger effects should not be discarded; (2) in semi-tauonic B decays both the SM and possible NP contributions are analytic throughout the relevant three-body decay phase space. On the contrary, the SM top decays are dominated by the $W$ pole. We have shown how existing experimental studies of tauonic top decays already constrain interesting LFU violating NP effects, provided the relevant NP degrees of freedom are light enough to be produced on-shell in top decays. Based on our findings we urge the experimental collaborations to extend their existing searches for charged scalars decaying to $\tau\nu$  in top decays to also target charged spin-1 bosons. 

In the case of heavy off-shell NP effects, the MFV-like scaling from the B decay studies generically implies prohibitively small contributions in top decay rates. In addition existing approaches to LFU violation in top decays are already becoming systematics limited at the precision of $\mathcal{O}(20\%)$. Therefore we have developed a novel strategy to probe possible LFU violating NP effects in top decays at the percent level or possibly beyond. Our approach exploits some boost invariant features of the $b$-jet energy spectra in top decays that are especially sensitive to contributions not respecting the SM two-body kinematics. Relying heavily on data-driven techniques to control the systematics related to $\tau$-tagging rates and backgrounds, we have found that sub-percent level LFU violating effects in top decays could be within reach. In particular, we have demonstrated the discriminating power of our proposed observable in Eq.~\eqref{eq:master} -- the (binned) LFU ratio of $b$-jet energy spectra -- on the example of a heavy off-shell vector mediator. Our systematic uncertainty and resulting NP sensitivity estimates are limited by our MC sample statistics which approximately correspond to the Run 2 LHC dataset.

We also note that this work is the first dedicated study of using the shapes in $t \bar t $ distributions in order to probe the LFU in top sector. In this paper, we have made conservative assumptions about the  $\tau$-tagging and mistagging rates. For technical reasons, our Monte Carlo sample is about an order of magnitude 
smaller than the data from the full HL LHC run, and our errors are dominated by MC statistics. Thus we are optimistic that with more dedicated searches using full detector simulations  and more refined techniques, such as those based on machine learning, one would expect significant improvement in potential reach.  

\begin{acknowledgments}

We would like to thank Roberto Franceschini, Dag Gillberg, and Beate Heinemann for useful discussions. JFK and DS are also grateful for the hospitality of the CERN Theory group where some of this work was completed. JFK acknowledges the financial support from the Slovenian Research Agency (research core funding No. P1-0035 and J1-8137).  DS  is  supported in part by the Natural  Sciences and Engineering Research Council of Canada (NSERC). 
\end{acknowledgments}

\appendix

\section{NP Effects on Peak Location of $E_b$ Distribution}
\label{app:peak}

In this Appendix we describe the effects of New Physics on the peak of the $E_b$ distribution. We can work in the rest frame of the decaying top where the $E_b$ distribution is an approximate $\delta$-function in the SM at an energy $E_b^*$ shown in Eq.~(\ref{eq:Estar}). The differential top width can be written in terms of $x_b = E_b/(2m_t)$ which ranges from 0 to 1. The SM, interference, and new physics squared contributions for a vector new physics model are given by:
\begin{eqnarray}
\frac{d\Gamma_{\rm SM}}{dx_b} &=& \frac{\alpha^2 m_t^5}{192\pi s_W^4}
\frac{x_b^2 (3-2x_b)}{\left(m_W^2 - m_t^2 (1-x_b) \right)^2 + m_W^2 \Gamma_W^2}\\
\frac{d\Gamma_{\rm int}}{dx_b} &=& \frac{\alpha \, \alpha_{NP} m_t^5}{96\pi s_W^2}
\frac{x_b^2 (3-2x_b)}{\left(m_\cV^2 - m_t^2 (1-x_b)\right) }
\frac{m_W^2-m_t^2(1-x_b)}{\left( \left(m_W^2 - m_t^2 (1-x_b) \right)^2 + m_W^2 \Gamma_W^2 \right)}\\
\frac{d\Gamma_{\rm NP}}{dx_b} &=& \frac{\alpha_{NP}^2 m_t^5}{192\pi}
\frac{x_b^2 (3-2x_b)}{\left(m_\cV^2 - m_t^2 (1-x_b) \right)^2},
\end{eqnarray}
where $s_W$ is the sine of the weak mixing angle, $\Gamma_W \approx 2$ GeV is the width of the $W$ boson, and $\alpha_{NP}$ and $m_\cV$ are the coupling squared and mass of the new vector, and these formulas assume $m_\cV > m_t$. From this we see that the SM contribution is largest for $x_b = (m_t^2-m_W^2)/m_t^2$, or equivalently, $E_b = E_b^*$. We can now see the qualitative effect of the new physics on the peak. The interference effect changes sign at $E_b = E_b^*$, so it will shift the peak to either higher or lower values depending on the sign relative to the SM. 
The non-interfering NP contributions peak at the largest $b$-jet energy $E^{\rm max}_b \sim
m_t/2$, shifting the overall $E_b$ distribution to higher values. 

 \begin{figure}
	\centering
	\includegraphics[width=.7\textwidth]{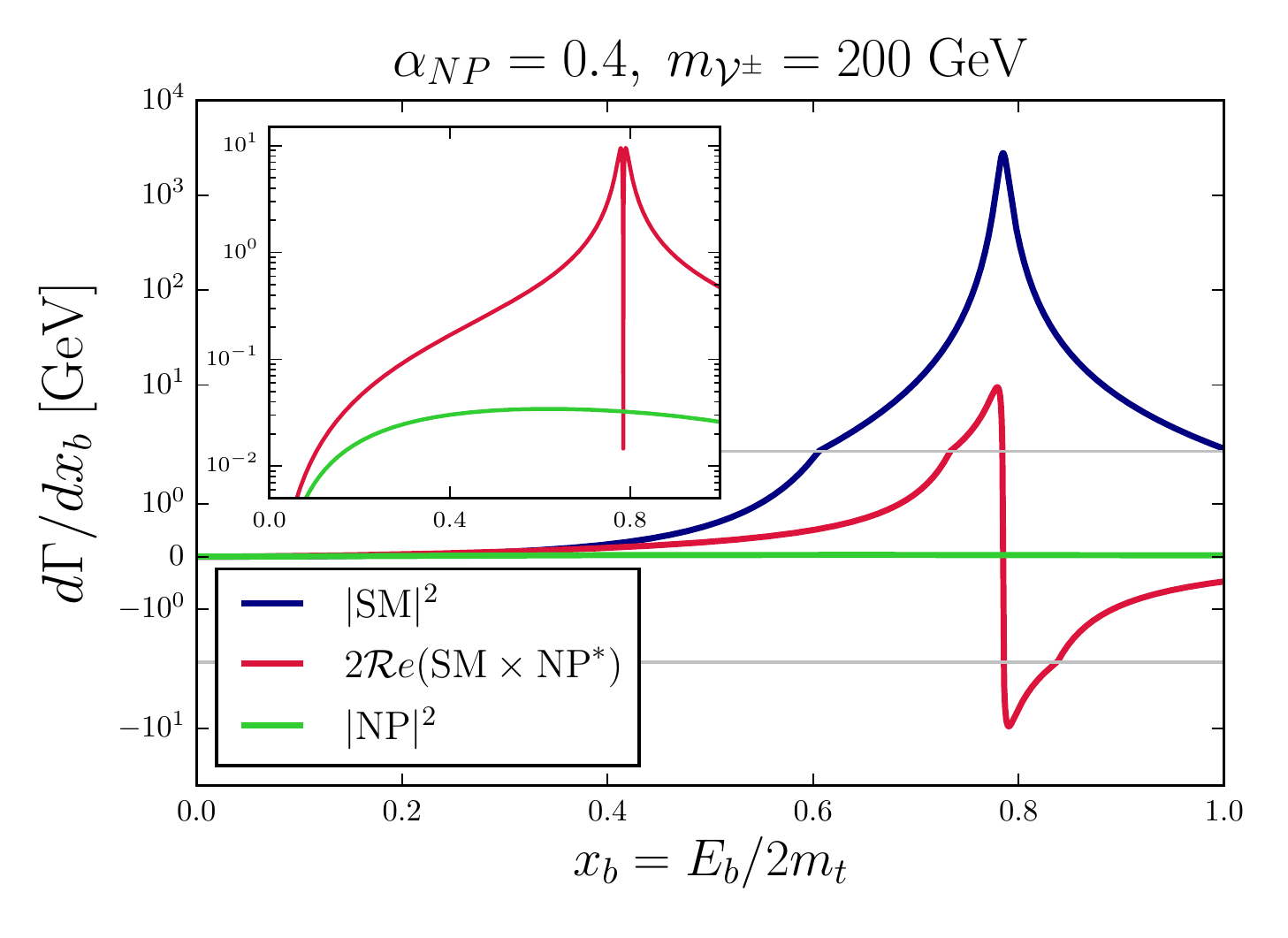}
	\caption{Differential decay rate of the SM (blue), the new physics squared (green), and the interference term (red) in the rest frame of the decaying
		top, assuming the vector model. The interference term does not shift the peak of the distribution, 
		however it does change the shape around it. The inset plot shows the absolute values new physics terms. The NP parameters that we have chosen is representative of the NP models considered in this work. Parenthetically 
		we point out that the small kinks on the main plot are not physical discontinuities, but rather a transition from the log scale to the linear one (the transition line is marked by light gray).   }
	\label{fig:analytic}
\end{figure}

The NP effects all turn out to be quantitatively extremely small relative to the SM near the region of the $E_b$ peak, namely,
\begin{equation}
\frac{d\Gamma_{\rm NP}}{dE_b} \ll 
\frac{d\Gamma_{\rm int}}{dE_b} \ll 
\frac{d\Gamma_{\rm SM}}{dE_b}, 
\;\;\; \text{for   }E_b \approx E_b^*.
\label{eq:scaling}
\end{equation}
This can be seen from the analytic expressions above. For the SM contribution 
$d\Gamma_{\rm SM}/dE_b \big\rvert_{E_b = E_b^*} \propto \Gamma_W^{-2}$ where $\Gamma_W \approx 2$ GeV is the width of the $W$ boson. In the case of a vector model, the interference term is zero for $E_b = E_b^*$, but near the peak, $d\Gamma_{\rm int}/dE_b \big\rvert_{E_b \approx E_b^*} \propto \Gamma_W^{-1}$, while for a scalar model the interference is negligible throughout the phase space. The pure new physics contribution to the top width is independent of the $W$ width: $d\Gamma_{\rm NP}/dE_b \big\rvert_{E_b = E_b^*} \propto \Gamma_W^{0}$, and for an off-shell mediator, it is also insensitive to width of the NP mediator at leading order. The other scales that make up the dimensions are the $W$, top, and new physics mass scales, which are all parametrically larger than the $\Gamma_W$. This scaling with the width of the $W$ confirms the hierarchy of Eq.~(\ref{eq:scaling}).

We illustrate these points explicitly in Fig.~\ref{fig:analytic} for a new physics parameter point comparable 
to those studied in this work, where we see that in the region of the peak, 
Eq.~(\ref{eq:scaling}) is well satisfied.

\renewcommand{\baselinestretch}{1.2} 
\small 
\bibliography{topbib}

\providecommand{\href}[2]{#2}\begingroup\raggedright\begin{thebibliography}{10}

\bibitem{Patrignani:2016xqp}
{\scshape Particle Data Group} collaboration, C.~Patrignani et~al.,
  \emph{{Review of Particle Physics}},
  \href{http://dx.doi.org/10.1088/1674-1137/40/10/100001}{\emph{Chin. Phys.}
  {\bf C40} (2016) 100001}.

\bibitem{ALEPH:2005ab}
{\scshape SLD Electroweak Group, DELPHI, ALEPH, SLD, SLD Heavy Flavour Group,
  OPAL, LEP Electroweak Working Group, L3} collaboration, S.~Schael et~al.,
  \emph{{Precision electroweak measurements on the $Z$ resonance}},
  \href{http://dx.doi.org/10.1016/j.physrep.2005.12.006}{\emph{Phys. Rept.}
  {\bf 427} (2006) 257--454}, [\href{https://arxiv.org/abs/hep-ex/0509008}{{\tt
  hep-ex/0509008}}].

\bibitem{Amhis:2016xyh}
{\scshape HFLAV} collaboration, Y.~Amhis et~al., \emph{{Averages of $b$-hadron,
  $c$-hadron, and $\tau$-lepton properties as of summer 2016}},
  \href{http://dx.doi.org/10.1140/epjc/s10052-017-5058-4}{\emph{Eur. Phys. J.}
  {\bf C77} (2017) 895}, [\href{https://arxiv.org/abs/1612.07233}{{\tt
  1612.07233}}].

\bibitem{Aaij:2014ora}
{\scshape LHCb} collaboration, R.~Aaij et~al., \emph{{Test of lepton
  universality using $B^{+}\rightarrow K^{+}\ell^{+}\ell^{-}$ decays}},
  \href{http://dx.doi.org/10.1103/PhysRevLett.113.151601}{\emph{Phys. Rev.
  Lett.} {\bf 113} (2014) 151601}, [\href{https://arxiv.org/abs/1406.6482}{{\tt
  1406.6482}}].

\bibitem{Aaij:2015oid}
{\scshape LHCb} collaboration, R.~Aaij et~al., \emph{{Angular analysis of the
  $B^{0} \to K^{*0} \mu^{+} \mu^{-}$ decay using 3 fb$^{-1}$ of integrated
  luminosity}}, \href{http://dx.doi.org/10.1007/JHEP02(2016)104}{\emph{JHEP}
  {\bf 02} (2016) 104}, [\href{https://arxiv.org/abs/1512.04442}{{\tt
  1512.04442}}].

\bibitem{Aaij:2017vbb}
{\scshape LHCb} collaboration, R.~Aaij et~al., \emph{{Test of lepton
  universality with $B^{0} \rightarrow K^{*0}\ell^{+}\ell^{-}$ decays}},
  \href{http://dx.doi.org/10.1007/JHEP08(2017)055}{\emph{JHEP} {\bf 08} (2017)
  055}, [\href{https://arxiv.org/abs/1705.05802}{{\tt 1705.05802}}].

\bibitem{Fajfer:2012jt}
S.~Fajfer, J.~F. Kamenik, I.~Nisandzic and J.~Zupan, \emph{{Implications of
  Lepton Flavor Universality Violations in B Decays}},
  \href{http://dx.doi.org/10.1103/PhysRevLett.109.161801}{\emph{Phys. Rev.
  Lett.} {\bf 109} (2012) 161801}, [\href{https://arxiv.org/abs/1206.1872}{{\tt
  1206.1872}}].

\bibitem{Freytsis:2015qca}
M.~Freytsis, Z.~Ligeti and J.~T. Ruderman, \emph{{Flavor models for $\bar{B}
  \to D^{(*)} \tau \bar{\nu}$}},
  \href{http://dx.doi.org/10.1103/PhysRevD.92.054018}{\emph{Phys. Rev.} {\bf
  D92} (2015) 054018}, [\href{https://arxiv.org/abs/1506.08896}{{\tt
  1506.08896}}].

\bibitem{Bardhan:2016uhr}
D.~Bardhan, P.~Byakti and D.~Ghosh, \emph{{A closer look at the R$_{D}$ and
  R$_{D^*}$ anomalies}},
  \href{http://dx.doi.org/10.1007/JHEP01(2017)125}{\emph{JHEP} {\bf 01} (2017)
  125}, [\href{https://arxiv.org/abs/1610.03038}{{\tt 1610.03038}}].

\bibitem{Celis:2016azn}
A.~Celis, M.~Jung, X.-Q. Li and A.~Pich, \emph{{Scalar contributions to $b\to c
  (u) \tau \nu$ transitions}},
  \href{http://dx.doi.org/10.1016/j.physletb.2017.05.037}{\emph{Phys. Lett.}
  {\bf B771} (2017) 168--179}, [\href{https://arxiv.org/abs/1612.07757}{{\tt
  1612.07757}}].

\bibitem{Bernlochner:2017jka}
F.~U. Bernlochner, Z.~Ligeti, M.~Papucci and D.~J. Robinson, \emph{{Combined
  analysis of semileptonic $B$ decays to $D$ and $D^*$: $R(D^{(*)})$,
  $|V_{cb}|$, and new physics}},
  \href{http://dx.doi.org/10.1103/PhysRevD.95.115008,
  10.1103/PhysRevD.97.059902}{\emph{Phys. Rev.} {\bf D95} (2017) 115008},
  [\href{https://arxiv.org/abs/1703.05330}{{\tt 1703.05330}}].

\bibitem{Hiller:2014yaa}
G.~Hiller and M.~Schmaltz, \emph{{$R_K$ and future $b \to s \ell \ell$ physics
  beyond the standard model opportunities}},
  \href{http://dx.doi.org/10.1103/PhysRevD.90.054014}{\emph{Phys. Rev.} {\bf
  D90} (2014) 054014}, [\href{https://arxiv.org/abs/1408.1627}{{\tt
  1408.1627}}].

\bibitem{Altmannshofer:2014rta}
W.~Altmannshofer and D.~M. Straub, \emph{{New physics in $b\rightarrow s$
  transitions after LHC run 1}},
  \href{http://dx.doi.org/10.1140/epjc/s10052-015-3602-7}{\emph{Eur. Phys. J.}
  {\bf C75} (2015) 382}, [\href{https://arxiv.org/abs/1411.3161}{{\tt
  1411.3161}}].

\bibitem{Altmannshofer:2015sma}
W.~Altmannshofer and D.~M. Straub, \emph{{Implications of $b\to s$
  measurements}},  in \emph{{Proceedings, 50th Rencontres de Moriond
  Electroweak Interactions and Unified Theories: La Thuile, Italy, March 14-21,
  2015}}, pp.~333--338, 2015.
\newblock \href{https://arxiv.org/abs/1503.06199}{{\tt 1503.06199}}.

\bibitem{Descotes-Genon:2015uva}
S.~Descotes-Genon, L.~Hofer, J.~Matias and J.~Virto, \emph{{Global analysis of
  $b\to s\ell\ell$ anomalies}},
  \href{http://dx.doi.org/10.1007/JHEP06(2016)092}{\emph{JHEP} {\bf 06} (2016)
  092}, [\href{https://arxiv.org/abs/1510.04239}{{\tt 1510.04239}}].

\bibitem{Capdevila:2017bsm}
B.~Capdevila, A.~Crivellin, S.~Descotes-Genon, J.~Matias and J.~Virto,
  \emph{{Patterns of New Physics in $b\to s\ell^+\ell^-$ transitions in the
  light of recent data}},
  \href{http://dx.doi.org/10.1007/JHEP01(2018)093}{\emph{JHEP} {\bf 01} (2018)
  093}, [\href{https://arxiv.org/abs/1704.05340}{{\tt 1704.05340}}].

\bibitem{Aad:2015dya}
{\scshape ATLAS} collaboration, G.~Aad et~al., \emph{{Measurements of the top
  quark branching ratios into channels with leptons and quarks with the ATLAS
  detector}}, \href{http://dx.doi.org/10.1103/PhysRevD.92.072005}{\emph{Phys.
  Rev.} {\bf D92} (2015) 072005}, [\href{https://arxiv.org/abs/1506.05074}{{\tt
  1506.05074}}].

\bibitem{CMS-PAS-HIG-12-052}
{\scshape CMS Collaboration} collaboration, \emph{{Updated search for a light
  charged Higgs boson in top quark decays in pp collisions at sqrt(s) = 7
  TeV}},  Tech. Rep. CMS-PAS-HIG-12-052, CERN, Geneva, 2012.

\bibitem{Becirevic:2016yqi}
D.~Bečirević, S.~Fajfer, N.~Košnik and O.~Sumensari, \emph{{Leptoquark model
  to explain the $B$-physics anomalies, $R_K$ and $R_D$}},
  \href{http://dx.doi.org/10.1103/PhysRevD.94.115021}{\emph{Phys. Rev.} {\bf
  D94} (2016) 115021}, [\href{https://arxiv.org/abs/1608.08501}{{\tt
  1608.08501}}].

\bibitem{Alpigiani:2017lpj}
C.~Alpigiani et~al., \emph{{Unitarity Triangle Analysis in the Standard Model
  and Beyond}},  in \emph{{5th Large Hadron Collider Physics Conference (LHCP
  2017) Shanghai, China, May 15-20, 2017}}, 2017.
\newblock \href{https://arxiv.org/abs/1710.09644}{{\tt 1710.09644}}.

\bibitem{Heister:2002ev}
{\scshape ALEPH} collaboration, A.~Heister et~al., \emph{{Search for charged
  Higgs bosons in $e^{+} e^{-}$ collisions at energies up to $\sqrt{s}$ =
  209-GeV}}, \href{http://dx.doi.org/10.1016/S0370-2693(02)02380-8}{\emph{Phys.
  Lett.} {\bf B543} (2002) 1--13},
  [\href{https://arxiv.org/abs/hep-ex/0207054}{{\tt hep-ex/0207054}}].

\bibitem{Achard:2003gt}
{\scshape L3} collaboration, P.~Achard et~al., \emph{{Search for charged Higgs
  bosons at LEP}},
  \href{http://dx.doi.org/10.1016/j.physletb.2003.09.057}{\emph{Phys. Lett.}
  {\bf B575} (2003) 208--220},
  [\href{https://arxiv.org/abs/hep-ex/0309056}{{\tt hep-ex/0309056}}].

\bibitem{Abbiendi:2008aa}
{\scshape OPAL} collaboration, G.~Abbiendi et~al., \emph{{Search for Charged
  Higgs Bosons in $e^+e^-$ Collisions at $\sqrt{s} = 189-209$ GeV}},
  \href{http://dx.doi.org/10.1140/epjc/s10052-012-2076-0}{\emph{Eur. Phys. J.}
  {\bf C72} (2012) 2076}, [\href{https://arxiv.org/abs/0812.0267}{{\tt
  0812.0267}}].

\bibitem{Greljo:2015mma}
A.~Greljo, G.~Isidori and D.~Marzocca, \emph{{On the breaking of Lepton Flavor
  Universality in B decays}},
  \href{http://dx.doi.org/10.1007/JHEP07(2015)142}{\emph{JHEP} {\bf 07} (2015)
  142}, [\href{https://arxiv.org/abs/1506.01705}{{\tt 1506.01705}}].

\bibitem{Boucenna:2016wpr}
S.~M. Boucenna, A.~Celis, J.~Fuentes-Martin, A.~Vicente and J.~Virto,
  \emph{{Non-abelian gauge extensions for B-decay anomalies}},
  \href{http://dx.doi.org/10.1016/j.physletb.2016.06.067}{\emph{Phys. Lett.}
  {\bf B760} (2016) 214--219}, [\href{https://arxiv.org/abs/1604.03088}{{\tt
  1604.03088}}].

\bibitem{Boucenna:2016qad}
S.~M. Boucenna, A.~Celis, J.~Fuentes-Martin, A.~Vicente and J.~Virto,
  \emph{{Phenomenology of an $SU(2) \times SU(2) \times U(1)$ model with
  lepton-flavour non-universality}},
  \href{http://dx.doi.org/10.1007/JHEP12(2016)059}{\emph{JHEP} {\bf 12} (2016)
  059}, [\href{https://arxiv.org/abs/1608.01349}{{\tt 1608.01349}}].

\bibitem{Celis:2012dk}
A.~Celis, M.~Jung, X.-Q. Li and A.~Pich, \emph{{Sensitivity to charged scalars
  in $\boldsymbol{B\to D^{(*)}\tau\nu_\tau}$ and
  $\boldsymbol{B\to\tau\nu_\tau}$ decays}},
  \href{http://dx.doi.org/10.1007/JHEP01(2013)054}{\emph{JHEP} {\bf 01} (2013)
  054}, [\href{https://arxiv.org/abs/1210.8443}{{\tt 1210.8443}}].

\bibitem{Crivellin:2013wna}
A.~Crivellin, A.~Kokulu and C.~Greub, \emph{{Flavor-phenomenology of
  two-Higgs-doublet models with generic Yukawa structure}},
  \href{http://dx.doi.org/10.1103/PhysRevD.87.094031}{\emph{Phys. Rev.} {\bf
  D87} (2013) 094031}, [\href{https://arxiv.org/abs/1303.5877}{{\tt
  1303.5877}}].

\bibitem{Dorsner:2016wpm}
I.~Doršner, S.~Fajfer, A.~Greljo, J.~F. Kamenik and N.~Košnik, \emph{{Physics
  of leptoquarks in precision experiments and at particle colliders}},
  \href{http://dx.doi.org/10.1016/j.physrep.2016.06.001}{\emph{Phys. Rept.}
  {\bf 641} (2016) 1--68}, [\href{https://arxiv.org/abs/1603.04993}{{\tt
  1603.04993}}].

\bibitem{Bauer:2015knc}
M.~Bauer and M.~Neubert, \emph{{Minimal Leptoquark Explanation for the
  R$_{D^{(*)}}$ , R$_K$ , and $(g-2)_g$ Anomalies}},
  \href{http://dx.doi.org/10.1103/PhysRevLett.116.141802}{\emph{Phys. Rev.
  Lett.} {\bf 116} (2016) 141802},
  [\href{https://arxiv.org/abs/1511.01900}{{\tt 1511.01900}}].

\bibitem{Aaboud:2016qeg}
{\scshape ATLAS} collaboration, M.~Aaboud et~al., \emph{{Search for scalar
  leptoquarks in pp collisions at $\sqrt{s}$ = 13 TeV with the ATLAS
  experiment}},
  \href{http://dx.doi.org/10.1088/1367-2630/18/9/093016}{\emph{New J. Phys.}
  {\bf 18} (2016) 093016}, [\href{https://arxiv.org/abs/1605.06035}{{\tt
  1605.06035}}].

\bibitem{CMS-PAS-B2G-16-028}
{\scshape CMS Collaboration} collaboration, \emph{{Search for third generation
  scalar leptoquarks decaying to a top quark and a tau lepton at sqrt(s) = 13
  TeV}},  Tech. Rep. CMS-PAS-B2G-16-028, CERN, Geneva, 2017.

\bibitem{Alwall:2014hca}
J.~Alwall, R.~Frederix, S.~Frixione, V.~Hirschi, F.~Maltoni, O.~Mattelaer
  et~al., \emph{{The automated computation of tree-level and next-to-leading
  order differential cross sections, and their matching to parton shower
  simulations}}, \href{http://dx.doi.org/10.1007/JHEP07(2014)079}{\emph{JHEP}
  {\bf 07} (2014) 079}, [\href{https://arxiv.org/abs/1405.0301}{{\tt
  1405.0301}}].

\bibitem{Alloul:2013bka}
A.~Alloul, N.~D. Christensen, C.~Degrande, C.~Duhr and B.~Fuks,
  \emph{{FeynRules 2.0 - A complete toolbox for tree-level phenomenology}},
  \href{http://dx.doi.org/10.1016/j.cpc.2014.04.012}{\emph{Comput. Phys.
  Commun.} {\bf 185} (2014) 2250--2300},
  [\href{https://arxiv.org/abs/1310.1921}{{\tt 1310.1921}}].

\bibitem{Sjostrand:2014zea}
T.~Sjöstrand, S.~Ask, J.~R. Christiansen, R.~Corke, N.~Desai, P.~Ilten et~al.,
  \emph{{An Introduction to PYTHIA 8.2}},
  \href{http://dx.doi.org/10.1016/j.cpc.2015.01.024}{\emph{Comput. Phys.
  Commun.} {\bf 191} (2015) 159--177},
  [\href{https://arxiv.org/abs/1410.3012}{{\tt 1410.3012}}].

\bibitem{deFavereau:2013fsa}
{\scshape DELPHES 3} collaboration, J.~de~Favereau, C.~Delaere, P.~Demin,
  A.~Giammanco, V.~Lemaître, A.~Mertens et~al., \emph{{DELPHES 3, A modular
  framework for fast simulation of a generic collider experiment}},
  \href{http://dx.doi.org/10.1007/JHEP02(2014)057}{\emph{JHEP} {\bf 02} (2014)
  057}, [\href{https://arxiv.org/abs/1307.6346}{{\tt 1307.6346}}].

\bibitem{CMS-PAS-SUS-17-003}
{\scshape CMS Collaboration} collaboration, \emph{{Search for pair production
  of tau sleptons in $\sqrt{s}=13~\mathrm{TeV}$ pp collisions in the
  all-hadronic final state}},  Tech. Rep. CMS-PAS-SUS-17-003, CERN, Geneva,
  2017.

\bibitem{Aaboud:2016dig}
{\scshape ATLAS} collaboration, M.~Aaboud et~al., \emph{{Search for charged
  Higgs bosons produced in association with a top quark and decaying via
  $H^{\pm} \rightarrow \tau\nu$ using $pp$ collision data recorded at $\sqrt{s}
  = 13$ TeV by the ATLAS detector}},
  \href{http://dx.doi.org/10.1016/j.physletb.2016.06.017}{\emph{Phys. Lett.}
  {\bf B759} (2016) 555--574}, [\href{https://arxiv.org/abs/1603.09203}{{\tt
  1603.09203}}].

\bibitem{ATLAS-CONF-2016-089}
{\scshape ATLAS Collaboration} collaboration, \emph{{Search for charged Higgs
  bosons in the $H^{\pm}\to tb$ decay channel in $pp$ collisions at
  $\sqrt{s}=13$ TeV using the ATLAS detector}},  Tech. Rep.
  ATLAS-CONF-2016-089, CERN, Geneva, Aug, 2016.

\bibitem{CMS-PAS-HIG-16-031}
{\scshape CMS Collaboration} collaboration, \emph{{Search for charged Higgs
  bosons with the $\mathrm{H}^{\scriptscriptstyle \pm}\rightarrow
  \tau^{\scriptscriptstyle \pm}\nu_{\tau}$ decay channel in the fully hadronic
  final state at $\sqrt{s} = 13~\mathrm{TeV}$}},  Tech. Rep.
  CMS-PAS-HIG-16-031, CERN, Geneva, 2016.

\bibitem{Aaboud:2018gjj}
{\scshape ATLAS} collaboration, M.~Aaboud et~al., \emph{{Search for charged
  Higgs bosons decaying via $H^{\pm} \to \tau^{\pm}\nu_{\tau}$ in the
  $\tau$+jets and $\tau$+lepton final states with 36 fb$^{-1}$ of $pp$
  collision data recorded at $\sqrt{s} = 13$ TeV with the ATLAS experiment}},
  {\emph{Submitted to: JHEP} (2018) },
  [\href{https://arxiv.org/abs/1807.07915}{{\tt 1807.07915}}].

\bibitem{Aaboud:2018cwk}
{\scshape ATLAS} collaboration, M.~Aaboud et~al., \emph{{Search for charged
  Higgs bosons decaying into top and bottom quarks at $\sqrt{s}$ = 13 TeV with
  the ATLAS detector}}, {\emph{Submitted to: JHEP} (2018) },
  [\href{https://arxiv.org/abs/1808.03599}{{\tt 1808.03599}}].

\bibitem{Agashe:2012bn}
K.~Agashe, R.~Franceschini and D.~Kim, \emph{{Simple “invariance” of
  two-body decay kinematics}},
  \href{http://dx.doi.org/10.1103/PhysRevD.88.057701}{\emph{Phys. Rev.} {\bf
  D88} (2013) 057701}, [\href{https://arxiv.org/abs/1209.0772}{{\tt
  1209.0772}}].

\bibitem{Agashe:2013eba}
K.~Agashe, R.~Franceschini and D.~Kim, \emph{{Using Energy Peaks to Measure New
  Particle Masses}},
  \href{http://dx.doi.org/10.1007/JHEP11(2014)059}{\emph{JHEP} {\bf 11} (2014)
  059}, [\href{https://arxiv.org/abs/1309.4776}{{\tt 1309.4776}}].

\bibitem{CMS-PAS-TOP-15-002}
{\scshape CMS Collaboration} collaboration, \emph{{Measurement of the top-quark
  mass from the b jet energy spectrum}},  Tech. Rep. CMS-PAS-TOP-15-002, CERN,
  Geneva, 2015.

\bibitem{Agashe:2015wwa}
K.~Agashe, R.~Franceschini, D.~Kim and K.~Wardlow, \emph{{Mass Measurement
  Using Energy Spectra in Three-body Decays}},
  \href{http://dx.doi.org/10.1007/JHEP05(2016)138}{\emph{JHEP} {\bf 05} (2016)
  138}, [\href{https://arxiv.org/abs/1503.03836}{{\tt 1503.03836}}].

\bibitem{Alwall:2011uj}
J.~Alwall, M.~Herquet, F.~Maltoni, O.~Mattelaer and T.~Stelzer, \emph{{MadGraph
  5 : Going Beyond}},
  \href{http://dx.doi.org/10.1007/JHEP06(2011)128}{\emph{JHEP} {\bf 06} (2011)
  128}, [\href{https://arxiv.org/abs/1106.0522}{{\tt 1106.0522}}].

\bibitem{Hoche:2006ph}
S.~Hoeche, F.~Krauss, N.~Lavesson, L.~Lonnblad, M.~Mangano, A.~Schalicke
  et~al., \emph{{Matching parton showers and matrix elements}},  in \emph{{HERA
  and the LHC: A Workshop on the implications of HERA for LHC physics:
  Proceedings Part A}}, pp.~288--289, 2005.
\newblock \href{https://arxiv.org/abs/hep-ph/0602031}{{\tt hep-ph/0602031}}.
\newblock \href{http://dx.doi.org/10.5170/CERN-2005-014.288}{DOI}.

\bibitem{Cacciari:2005hq}
M.~Cacciari and G.~P. Salam, \emph{{Dispelling the $N^{3}$ myth for the $k_t$
  jet-finder}},
  \href{http://dx.doi.org/10.1016/j.physletb.2006.08.037}{\emph{Phys. Lett.}
  {\bf B641} (2006) 57--61}, [\href{https://arxiv.org/abs/hep-ph/0512210}{{\tt
  hep-ph/0512210}}].

\bibitem{Cacciari:2011ma}
M.~Cacciari, G.~P. Salam and G.~Soyez, \emph{{FastJet User Manual}},
  \href{http://dx.doi.org/10.1140/epjc/s10052-012-1896-2}{\emph{Eur. Phys. J.}
  {\bf C72} (2012) 1896}, [\href{https://arxiv.org/abs/1111.6097}{{\tt
  1111.6097}}].

\bibitem{Cacciari:2008gp}
M.~Cacciari, G.~P. Salam and G.~Soyez, \emph{{The Anti-k(t) jet clustering
  algorithm}},
  \href{http://dx.doi.org/10.1088/1126-6708/2008/04/063}{\emph{JHEP} {\bf 04}
  (2008) 063}, [\href{https://arxiv.org/abs/0802.1189}{{\tt 0802.1189}}].

\bibitem{Katz:2010iq}
A.~Katz, M.~Son and B.~Tweedie, \emph{{Ditau-Jet Tagging and Boosted Higgses
  from a Multi-TeV Resonance}},
  \href{http://dx.doi.org/10.1103/PhysRevD.83.114033}{\emph{Phys. Rev.} {\bf
  D83} (2011) 114033}, [\href{https://arxiv.org/abs/1011.4523}{{\tt
  1011.4523}}].

\bibitem{Khachatryan:2015dfa}
{\scshape CMS} collaboration, V.~Khachatryan et~al., \emph{{Reconstruction and
  identification of $\tau$ lepton decays to hadrons and $\nu_\tau$ at CMS}},
  \href{http://dx.doi.org/10.1088/1748-0221/11/01/P01019}{\emph{JINST} {\bf 11}
  (2016) P01019}, [\href{https://arxiv.org/abs/1510.07488}{{\tt 1510.07488}}].

\bibitem{ATL-PHYS-PUB-2015-045}
\emph{{Reconstruction, Energy Calibration, and Identification of Hadronically
  Decaying Tau Leptons in the ATLAS Experiment for Run-2 of the LHC}},  Tech.
  Rep. ATL-PHYS-PUB-2015-045, CERN, Geneva, Nov, 2015.

\bibitem{Flechl:2017bse}
{\scshape ATLAS, CMS} collaboration, M.~Flechl, \emph{{Identification and
  energy calibration of hadronic tau lepton decays at the LHC}},  in \emph{{5th
  Large Hadron Collider Physics Conference (LHCP 2017) Shanghai, China, May
  15-20, 2017}}, 2017.
\newblock \href{https://arxiv.org/abs/1709.01351}{{\tt 1709.01351}}.

\bibitem{Khachatryan:2014loa}
{\scshape CMS} collaboration, V.~Khachatryan et~al., \emph{{Measurement of the
  $t \bar t$ production cross section in $pp$ collisions at $\sqrt s = 8$ TeV
  in dilepton final states containing one $\tau$ lepton}},
  \href{http://dx.doi.org/10.1016/j.physletb.2014.10.032}{\emph{Phys. Lett.}
  {\bf B739} (2014) 23--43}, [\href{https://arxiv.org/abs/1407.6643}{{\tt
  1407.6643}}].

\bibitem{Aad:2015tin}
{\scshape ATLAS} collaboration, G.~Aad et~al., \emph{{Search for direct top
  squark pair production in final states with two tau leptons in pp collisions
  at $\sqrt{s}=8$ TeV with the ATLAS detector}},
  \href{http://dx.doi.org/10.1140/epjc/s10052-016-3897-z}{\emph{Eur. Phys. J.}
  {\bf C76} (2016) 81}, [\href{https://arxiv.org/abs/1509.04976}{{\tt
  1509.04976}}].

\bibitem{Czakon:2013goa}
M.~Czakon, P.~Fiedler and A.~Mitov, \emph{{Total Top-Quark Pair-Production
  Cross Section at Hadron Colliders Through $O(\alpha^4_S)$}},
  \href{http://dx.doi.org/10.1103/PhysRevLett.110.252004}{\emph{Phys. Rev.
  Lett.} {\bf 110} (2013) 252004}, [\href{https://arxiv.org/abs/1303.6254}{{\tt
  1303.6254}}].

\bibitem{Czakon:2011xx}
M.~Czakon and A.~Mitov, \emph{{Top++: A Program for the Calculation of the
  Top-Pair Cross-Section at Hadron Colliders}},
  \href{http://dx.doi.org/10.1016/j.cpc.2014.06.021}{\emph{Comput. Phys.
  Commun.} {\bf 185} (2014) 2930}, [\href{https://arxiv.org/abs/1112.5675}{{\tt
  1112.5675}}].

\bibitem{Frederix:2011qg}
R.~Frederix, S.~Frixione, V.~Hirschi, F.~Maltoni, R.~Pittau and P.~Torrielli,
  \emph{{W and $Z/\gamma*$ boson production in association with a
  bottom-antibottom pair}},
  \href{http://dx.doi.org/10.1007/JHEP09(2011)061}{\emph{JHEP} {\bf 09} (2011)
  061}, [\href{https://arxiv.org/abs/1106.6019}{{\tt 1106.6019}}].

\end{thebibliography}\endgroup
\bibliographystyle{JHEP}

\end{document}